\documentclass[fleqn,usenatbib]{mnras}

\usepackage[T1]{fontenc}
\usepackage{ae,aecompl}

\usepackage{graphicx}    
\usepackage{amsmath}    
\usepackage{amssymb}    
\usepackage{stfloats}
\usepackage{pdflscape} 

\title[DES Y3 Galaxy Morphology Classification Catalogue]{Galaxy Morphological Classification Catalogue of the Dark Energy Survey Year 3 data with Convolutional Neural Networks}
\author[Ting-Yun Cheng et al.]{Ting-Yun Cheng,$^{1,2}$\thanks{E-mail:ting-yun.cheng@durham.ac.uk}
Christopher J. Conselice,$^{2,3}$
Alfonso Arag\'on-Salamanca,$^{2}$
\newauthor
M.~Aguena,$^{4,5}$
S.~Allam,$^{6}$
F.~Andrade-Oliveira,$^{7,5}$
J.~Annis,$^{6}$
A.~F.~L.~Bluck,$^{8, 9}$
\newauthor
D.~Brooks,$^{10}$
D.~L.~Burke,$^{11, 12}$
M.~Carrasco~Kind,$^{13, 14}$
J.~Carretero,$^{15}$
A.~Choi,$^{16}$
\newauthor
M.~Costanzi,$^{17, 18, 19}$
L.~N.~da Costa,$^{5, 20}$
M.~E.~S.~Pereira,$^{21}$
J.~De~Vicente,$^{22}$
\newauthor
H.~T.~Diehl,$^{6}$
A.~Drlica-Wagner,$^{23, 6, 24}$
K.~Eckert,$^{25}$
S.~Everett,$^{26}$
A.~E.~Evrard,$^{27, 21}$
\newauthor
I.~Ferrero,$^{28}$
P.~Fosalba,$^{29, 30}$
J.~Frieman,$^{6, 24}$
J.~Garc\'ia-Bellido,$^{31}$
D.~W.~Gerdes,$^{27, 21}$
\newauthor
T.~Giannantonio,$^{32, 9}$
D.~Gruen,$^{33, 11, 12}$
R.~A.~Gruendl,$^{13, 14}$
J.~Gschwend,$^{5, 20}$
\newauthor
G.~Gutierrez,$^{6}$
S.~R.~Hinton,$^{34}$
D.~L.~Hollowood,$^{26}$
K.~Honscheid,$^{16, 35}$
D.~J.~James,$^{36}$
\newauthor
E.~Krause,$^{37}$
K.~Kuehn,$^{38, 39}$
N.~Kuropatkin,$^{6}$
O.~Lahav,$^{10}$
M.~A.~G.~Maia,$^{5, 20}$
\newauthor
M.~March,$^{25}$
F.~Menanteau,$^{13, 14}$
R.~Miquel,$^{40, 15}$
R.~Morgan,$^{41}$
F.~Paz-Chinch\'{o}n,$^{13, 32}$
\newauthor
A.~Pieres,$^{5, 20}$
A.~A.~Plazas~Malag\'on,$^{42}$
A.~Roodman,$^{11, 12}$
E.~Sanchez,$^{22}$
V.~Scarpine,$^{6}$
\newauthor
S.~Serrano,$^{29, 30}$
I.~Sevilla-Noarbe,$^{22}$
M.~Smith,$^{43}$
M.~Soares-Santos,$^{21}$
E.~Suchyta,$^{44}$
\newauthor
M.~E.~C.~Swanson,$^{13}$
G.~Tarle,$^{21}$
D.~Thomas,$^{45}$
C.~To $^{33, 11, 12}$
\\
\\
\parbox{\textwidth}{\centering \textsc{\Large  } \\ \centering \textit{Author affiliations are listed at the end of this paper} }
}

\date{Accepted XXX. Received YYY; in original form ZZZ}

\pubyear{2020}

\begin{document}
\label{firstpage}
\pagerange{\pageref{firstpage}--\pageref{lastpage}}
\maketitle

\begin{abstract}
We present in this paper one of the largest galaxy morphological classification catalogues to date, including over 20 million of galaxies, using the Dark Energy Survey (DES) Year 3 data based on Convolutional Neural Networks (CNN). Monochromatic $i$-band DES images with linear, logarithmic, and gradient scales, matched with debiased visual classifications from the Galaxy Zoo 1 (GZ1) catalogue, are used to train our CNN models. With a training set including bright galaxies ($16\le{i}<18$) at low redshift ($z<0.25$), we furthermore investigate the limit of the accuracy of our predictions applied to galaxies at fainter magnitude and at higher redshifts. Our final catalogue covers magnitudes $16\le{i}<21$, and redshifts $z<1.0$, and provides predicted probabilities to two galaxy types -- Ellipticals and Spirals (disk galaxies). Our CNN classifications reveal an accuracy of over 99\% for bright galaxies when comparing with the GZ1 classifications ($i<18$). For fainter galaxies, the visual classification carried out by three of the co-authors shows that the CNN classifier correctly categorises disky galaxies with rounder and blurred features, which humans often incorrectly visually classify as Ellipticals. As a part of the validation, we carry out one of the largest examination of non-parametric methods, including $\sim$100,000 galaxies with the same coverage of magnitude and redshift as the training set from our catalogue. We find that the Gini coefficient is the best single parameter discriminator between Ellipticals and Spirals for this data set.

\end{abstract}


\section{Introduction}
\label{ch4_sec_intro}

Galaxy morphology is linked to the stellar populations of galaxies, providing essential clues to their formation history and evolution. Visual morphological classification was pioneered by \citet{Hubble1926}. His system initially had two broad galaxy morphological types, early-type galaxies (ETGs) and late type galaxies (LTGs), based on their appearance in optical light. These two broad categories connect galaxy morphology with a variety of stellar and structural properties. For instance, ETGs are dominated by older stellar populations and have no spiral structure, while LTGs usually contain a younger stellar population and often have spiral arms. These differences in stellar properties indicate that galaxies with different morphologies are at different evolutionary stages and evolution paths. Therefore, the availability of galaxy morphologies for very large samples is of great importance when studying the formation and evolution of galaxies.

Conventionally, visual assessment is the main method of galaxy morphological classification \citep[e.g.][]{deVaucouleurs1959, deVaucouleurs1964, Sandage1961, Fukugita2007, Nair2010, Baillard2011}. Since around 2000 there has been a significant growth in the size of imaging data sets and increasingly complex ones from e.g., the Hubble Space Telescopes.  Due to this and the development of computational capacity, non-parametric methods were developed such as the {\it CAS system} (Concentration, Asymmetry, and Smoothness/Clumpiness), the Gini coefficient, and the M20 parameter \citep{Conselice2003, Abraham2003, Lotz2004, Law2007}.  There are good indications that these parameters, which make no assumptions about the galaxy, are largely free from subjective biases. However, even these computational methods become challenging to apply when the astronomical data become too large and we have to use Big Data techniques and machine learning.   We are now in this era with the extensive imaging now provided by the Dark Energy Survey \footnote{https://www.darkenergysurvey.org/}\citep[DES;][]{Abbott2018} which has imaged over hundreds of millions of galaxies.   This is just the first of many upcoming imaging surveys which will be carried out in the coming decade, including from the Vera Rubin Observatory and Euclid.

Another successful approach for carrying out large scale morphological analyses are the ``Galaxy Zoo'' projects \citep{Lintott2008, Lintott2011, Willett2013}, designed initially for classifying galaxies in the Sloan Digital Sky Survey (SDSS).  This Galaxy Zoo is such that amateurs classify galaxies by answering a series of questions based on galaxy images through an online interface.   Studies resulting from Galaxy Zoo show the usefulness of the input from non-professionals in morphological classification of galaxies.  With many volunteers this process accelerates the classification procedure by including the general public rather than limiting these efforts to experts. However, the size of astronomical data generated by large scale surveys such as DES and future surveys such as the Vera Rubin Observatory Legacy Survey of Space and Time and the Euclid Space Telescope has increased to the stage that it would take on the order of > 100 years to classify with Galaxy Zoo. Therefore, machine learning techniques are critical for analysing large-scale astronomical data set, such as galaxy images.

The concept of machine learning in computational science started from \citet{Fukushima1975,Fukushima1980} and \citet{Fukushima1983}. For the past decades, machine learning techniques have been widely used in astronomical studies, such as star-galaxy separation\citep[e.g.][]{Odewahn1992, Weir1995}, strong lensing identification \citep[e.g.][]{Jacobs2017, Petrillo2017, Lanusse2018, Cheng2020b}, finding galaxy mergers \citep[e.g.][]{Bottrell2019, Ferreira2020}, among many other applications.   Since these early papers the computational capability and machine learning methodologies have made a remarkable improvement and machine learning is becoming a standard tool in astronomical investigations.
Specifically within galaxy morphological classifications, there are a slew of studies applying different supervised machine learning approaches \citep[e.g.][]{Huertas-Company2008, Huertas-Company2009, Huertas-Company2011, Shamir2009, Polsterer2012, Sreejith2018, Dubath2011, Miller2017, Beck2018}, neural networks \citep[e.g.][]{Maehoenen1995, Naim1995, Lahav1996, Ball2004, Banerji2010}, and Convolutional Neural Networks \citep[CNN; e.g.][]{Dieleman2015, Huertas-Company2015, Huertas-Company2018, Dominguez-Sanchez2018, Walmsley2020, Cheng2020a, Hausen2020, Ghosh2020}.

In this new study, we apply the CNN set up and calibration investigated and assembled in \citet[][hereafter, C20]{Cheng2020a} to predict probabilities of binary galaxy morphological classification for the DES Year 3 GOLD data \citep[hereafter, DES Y3 data;][]{Sevilla-Noarbe2021}. This project allows us to build one of the largest catalogues of galaxy morphological classification to date which includes $\sim$20 million resolved galaxies, along with the companion DES catalogue produced in \citet{Vega-Ferrero2020}. Both studies use the Dark Energy Survey imaging data; however, there is only a $\sim60$ per cent overlapping in samples between the two due to different initial sample selection criteria applied. Their approach involves simulating bright galaxies to a fainter magnitude for training, and uses multi-band images, while we use single band images of bright galaxies and include linear, gradient, and logarithmic images to emphasise different shapes and light distribution of galaxies for training.   Our paper is therefore based on single-band apparent morphologies, similar to how visual estimates have been carried out for the past 100 years. The two works use different methodologies and training setups as well. The comparison of the two studies is ongoing and will provide a solid validation in morphological classification of the overlap samples using the different approaches. This will give an insight for future deep learning applications in galaxy morphological classification, but this type of detail is beyond the scope of this catalog paper.   Since it is a large amount of work to compare the two catalogues, which includes more than 20 million galaxies each, a detailed comparison of the two studies is separate from this paper.

The arrangement for this paper is as follows. The data sets are described in Section~\ref{ch4_sec_data}, and we introduce the CNN used in the paper in Section~\ref{ch4_sec_cnn}. Other catalogues used for validating our CNN predictions are introduced in Section~\ref{ch4_sec_cata4comp}. The examination of the predictions are shown in Section~\ref{ch4_sec_validation}, while the content of our classification catalogue is presented in Section~\ref{ch4_sec_y3catalog}. Finally, we summarise this study in Section~\ref{ch4_sec_conclusion}.

\section{Data Sets}
\label{ch4_sec_data}

The Dark Energy Survey \citep[DES;][]{DES2005, DES2016} is a wide-field optical imaging survey covering 5000 square degrees \citep[$\sim$1/8 sky;][]{Neilsen2019} which partially overlaps with the survey area of the Sloan Digital Sky Survey (SDSS), but has a better imaging quality and deeper depth than the SDSS images. The Dark Energy Camera \citep[DECam;][]{Flaugher2015} is used in DES which has a high quantum efficiency in the red wavebands ($>90$ per cent from $\sim$650 to $\sim$900 nm), and gives images with a good image quality for imaging observations of  distant objects compared with previous surveys with the spatial resolution of 0.263 arcsec per pixel and the single epoch depth of $i=22.51$ \citep{Abbott2018}. An individual DES survey exposure has more than 500M pixels. Each coadd (tile) image covers 1/2 square degrees and has a size of 10,000 $\times$10,000 pixels. To create the galaxy stamps for this study, we follow the guideline in C20 (details in Section~\ref{ch4_sec_preprocessing}) and apply the same pre-processing procedure used in the paper to both the training set (Section~\ref{ch4_sec_traindata}) and the DES Y3 data (Section~\ref{ch4_sec_y3data}). In the next subsections we give an overview of how we prepare our data for analysis from the DES imaging.

\subsection{Pre-processing}
\label{ch4_sec_preprocessing}

\begin{figure*}{}
\begin{center}
\graphicspath{}
	\includegraphics[width=2\columnwidth]{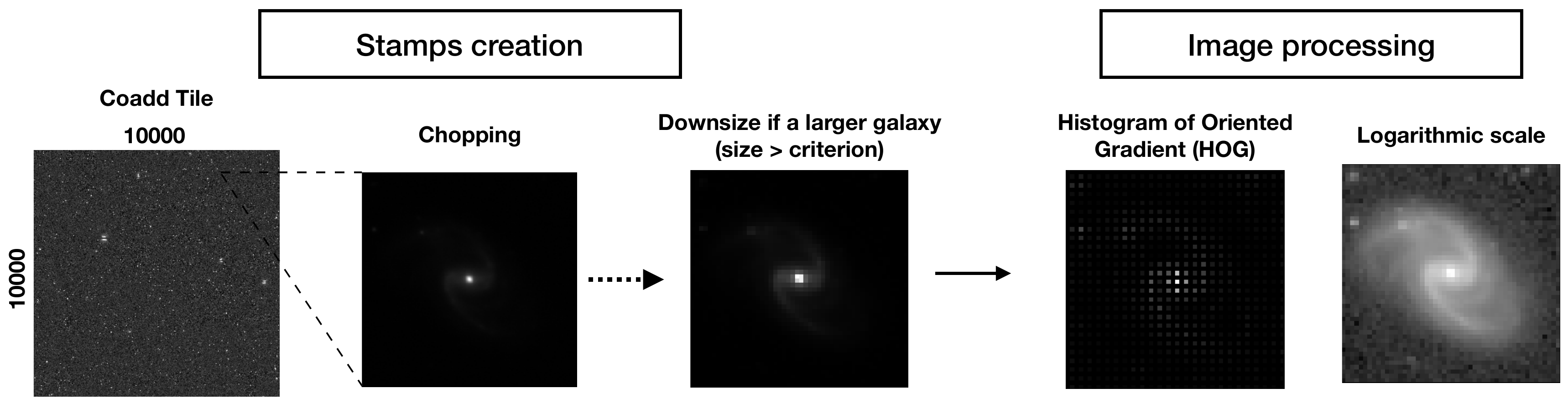}
   	\caption{Pre-processing procedure pipeline which is used to prepare our galaxy sample for analysis (see details in Section~\ref{ch4_sec_preprocessing}).  This shows the chopping, resizing processes (if needed), and imaging processing we utilise,  including histogram of oriented gradients (HOD) and using a logarithmic scaling as input.  }
    	\label{ch4_fig_preprocessing}
\end{center}
\end{figure*}

The data preparation we use closely follows the procedure described in C20. There are two main parts of the data preparation: (1) stamp creation and (2) image processing. Fig.~\ref{ch4_fig_preprocessing} shows the pre-processing procedure used in this study. Using the DES GOLD catalogues, we cut the original coadd images, which have a size of 10000$\times$10000 pixels, into many different `postage stamp' images -- creating millions of galaxy stamps with sizes of 50$\times$50 pixels (approximately $13^{ \prime \prime  }\times13^{ \prime \prime  }$). When a galaxy size, as given in the DES catalogue, is larger than the size threshold (30$\times$30 pixels), a larger 200$\times$200 pixel stamp is cut from the images, and then re-sampled to produce a 50$\times$50 pixel image by calculating the mean value in 4$\times$4 pixel blocks. This is done for a very small fraction of the galaxy sample since over 99\% of all DES galaxies are smaller than 25$\times$25 pixels. Additionally, when creating stamps for the training set, each image is rotated by different angles to increase the number of training images (see Section~\ref{ch4_sec_traindata}).

In the second step, we create two extra images which are both included in training our CNN models. One is an image with gradient features that we obtain by a feature extraction technique called the Histogram of Oriented Gradient \citep[HOG;][]{Dalal2005}. The HOG, as a feature extractor, is a well-known technique within pattern recognition studies, e.g. human detection, face recognition, and handwriting recognition \citep[e.g.,][etc]{Dalal2005, Shu2011, Kamble2015}. In astronomy, it has already been used in a few of studies such as spectral lines observation \citep{Soler2019}, gravitational lensing detection \citep[]{Avestruz2019}, and galaxy morphological classification such as our previous work \citep{Cheng2020a}.

The key feature of HOG is to characterise the local appearance and the shape of objects based on local intensity gradients \citep{Dalal2005}. This technique calculates the gradients of the horizontal (x) and vertical (y) direction of stamps. The magnitude and orientation of the gradient are calculated as below,
\begin{equation}
    	\left| G \right| =\sqrt { { G }_{ x }^{ 2 }+{ G }_{ y }^{ 2 } }, \\ \theta =\arctan { \left( \frac { { G }_{ y } }{ { G }_{ x } }  \right)  }
	\label{ch4_eq:hog}
\end{equation}
where $\left| G \right|$ is the gradient magnitude of each pixel, ${ G }_{ x }$ is the gradient magnitude measured in x-direction, ${ G }_{ y }$ is measured in the y-direction, and $\theta$ is the orientation of the gradient for each pixel in the images. It then measures the contribution of gradients from each pixel in the cell with a size of 2 by 2 pixels, and describes these using a histogram of different orientation angles. We rescale the HOG output images so that their pixel values are between 0 and 1 (hereafter, HOG images), and use them as one of the inputs to train our CNN models.

In addition to the HOG images, the other input we use is the image itself within a logarithmic scale (hereafter, log images). In C20, we tested the impact of using log images to train the CNN algorithms. We show in C20 that the improvement rate by using log images is positive, but decreased when the number of training data is increased. Therefore, there might not be a significant improvement provided by log images in our case. However, in order to completely consider different significant features in our images, we decide to include the log images with rescaled pixel values betwen 0 and 1 when training the final CNN models for the task of catalogue construction.

\subsection{Training Data - DES Y1 Data}
\label{ch4_sec_traindata}
The training data used throughout are described in C20, which is the subset of the first year DES GOLD data (DES Y1 data), the DES observation of SDSS stripe 82 \citep{Drlica-Wagner2018} and matched with the visual binary classifications from the Galaxy Zoo 1 project \citep[hereafter, GZ1\footnote{https://data.galaxyzoo.org/\label{GZ1web}}, ][Section~\ref{ch4_sec_gz_cata}]{Lintott2008, Lintott2011}. In this paper, the morphological classification catalogue is built based upon monochromatic $i$-band images only, due to the limitation of our computational resources and the cost of computational time to generate the pre-processed images and the memory storage of the enormous size of the DES Y3 data.

We directly used the visual classification (with over 80\% vote agreements) provided in \citet[][`Flags' from their Table 2; `{\it morphological flags}' hereafter]{Lintott2011}, giving us 2,862 galaxies with classification labels in total to train our machine. The instrinsic ratio between the number of Spirals and Ellipticals in this catalog is $\sim$3. The magnitude range of the overlap data ranges from $16\le{i}<18$, and their redshifts are all at $z<0.25$. However, in C20, we show how to correct the labels for $\sim2.5\%$ of our sample galaxies that are found to be mislabelled in GZ1 by comparing to DES data, which has a better resolution and deeper depth than the SDSS data; in which, a few galaxies are mislabelled due to the debias process carried out in GZ1 for creating the {\it morphological flags} (details in Section~\ref{ch4_sec_gz_cata}). Additionally, the $\sim0.56\%$ of galaxies that we cannot confirm the classification  for according to our test in C20 are excluded from our final training set. The final number of galaxies in our initial  training set is 2,846, with a ratio of $\sim$3 between the number of Spirals and Ellipticals.

The training set is prepared following the pipeline shown in Fig.~\ref{ch4_fig_preprocessing}. Considering we have a limited amount of labelled data, to prevent from overfitting during the training process, an extra process of rotating images is performed to increase the number of the training data. An extra amount of Gaussian noise (mean=0, variance=1E-8) is also added, which is negligible towards causing any impact to the signal-to-noise ratio, the visual appearance and the structure of galaxies, but shows a detectable change of pixel values \citep{Dieleman2015, Huertas-Company2015}. We do this to increase the variation of pixel values while increasing the number of training sets. Finally, we retain the balance between the number of elliptical (E) and spiral galaxies (S) in the training set which is proved to be an important factor in C20. This rotational operation increases the number of data for training purposes (including training and validation) to 54,133 galaxy stamps with the ratio of number of types held to $E/S\sim1$.

\subsection{The Galaxy Zoo 1 catalogue (GZ1)}
\label{ch4_sec_gz_cata}
The Galaxy Zoo projects are amongst the most successful attempts using citizen science to obtain large numbers of galaxy morphological classifications. A set of questions are asked to the volunteers for each galaxy image. Based on the answers from the volunteers, the GZ1 statistically provides the morphological classification of $\sim900,000$ galaxies. Of these, $\sim670,000$ galaxies with spectroscopic redshifts have been bias corrected \citep{Bamford2009}.

In this study, we use three main pieces of classification information from GZ1: {\it raw votes}, {\it debiased votes}, and {\it morphological flags}. The {\it raw votes} are the likelihood calculated directly from the volunteers’ votes for each image. The {\it debiased votes} and {\it morphological flags} are derived after applying bias corrections based upon different assumed Ellipticals/Spirals ratio \citep[hereafter, E/S ratio;][]{Lintott2011,Bamford2009}.

In GZ1, a correction factor is necessary to account for a classification bias that depends on the apparent brightness and size of each galaxy. For example, when viewing a spiral galaxy at higher redshift, its decreasing apparent brightness and size makes it more difficult to appreciate morphological details such as spiral arms, resulting in an increased likelihood of it being classified as an elliptical galaxy. The corrections needed to account for this bias are calculated by assuming that the morphological mix does not evolve significantly in the narrow redshift range covered by GZ1 \citep{Bamford2009}. This assumption has been shown to be a reasonable one \citep{Conselice2005}.

In order to perform this bias correction, GZ1 use two different values for the E/S ratio, one to obtain the {\it morphological flags} and a different one to estimate the {\it debiased votes}. The {\it morphological flags} provided by \citet{Lintott2011} are determined using the E/S values that only take into account the classifications with at least a 0.8 morphological vote fraction.  On the other hand, the {\it debiased votes} provided by GZ1 are based on E/S ratios that use the raw likelihood.

In C20, we note that some galaxies with less accurate {\it morphological flags} after the bias correction from the GZ1 still showed a questionable label when comparing with our CNN predictions. Therefore, through repeated tests of our CNN and visual assessment, we correct the labels for $\sim2.5\%$ of our sample galaxies, and excluded $\sim0.56\%$ galaxies that we cannot confirm the classification for based on our tests in C20. The corrected labels in C20 are shown to better corresponds to the classification based on the {\it debiased votes} in \citet[][`Debiased votes' in their Table 2]{Lintott2011} which is debiased based on the E/S ratio using the likelihoods directly \citep{Bamford2009}. The {\it debiased vote}, as stated on the website of the GZ1,  corrects well the bias existed in GZ1 visual assessment, and provides a more accurate classification than the {\it morphological flags}. Therefore, although our CNN model is trained with the {\textbf{\textit{corrected}}} {\it morphological flags} based on DES imaging data, we validate our CNN predictions with the classification based on {\it debiased votes} (Section~\ref{ch4_sec_compgz}).

\subsection{DES Year 3 Data}
\label{ch4_sec_y3data}
The data used to build the catalogue presented in this work is from the DES Y3 GOLD catalog \citep{Sevilla-Noarbe2021}. We initially use the images that are selected with the flags shown in Table~\ref{ch4_tab_flag} and within a magnitude range $16\le{i}\le22$. The top two flags guarantee that astronomical objects selected using these flags are most likely to be galaxies. This is such that the galaxy sample, as defined by these flags, has a rate of contamination of point sources less than $\sim$2\% at fainter magnitudes, as derived by comparing with the HSC-SSP DR2 catalog \citep{Aihara2018}. The bottom four flags ensure objects have a consistency with the Y3 GOLD footprint, denote quality selection for clean samples, and are used to select the data with a reliable {\sc SExtractor} \citep{Bertin1996} analysis.

This selection provides over 50 million galaxies for the initial task, with the redshift distribution of the selected data peak at $z\sim0.4$ with over $99.9\%$ of the galaxies at $z\le1.2$. The number of galaxies in each magnitude bin increases exponentially when going fainter. A pre-processing procedure described in Section~\ref{ch4_sec_preprocessing} is also applied to the selected data.

The selected data is separated into six magnitude bins from $i=16$ to $i=22$ for further analysis (Section~\ref{ch4_sec_validation}). After an examination carried out in Section~\ref{ch4_sec_compvis}, our final catalogue includes over 20 million galaxies with the magnitude range of $16\le{i}<21$.

The galaxies in the final catalogue have a wider range of magnitudes and redshifts than those in the training set - the training set galaxies are, typically, brighter and have lower redshifts (Section~\ref{ch4_sec_traindata}). Therefore, in our study, we also investigate how the confidence of our CNN predictions might be impacted when apply to galaxies at fainter magnitude and higher redshifts (Section~\ref{ch4_sec_confidence_level}). Alternatively, approaches such as using simulated data can help to build a training set that reaches fainter magnitudes or higher redshifts, e.g., the companion DES morphological catalogue presented in \citet{Vega-Ferrero2020}.
\begin{table}
	\centering
	\begin{tabular}{lc}
		\hline
		\multicolumn{1}{c}{Selection Flags} & {}\\
		\hline\hline
		\multicolumn{1}{l}{EXTENDED\_CLASS\_COADD} & {= 3} \\
		\multicolumn{1}{l}{EXTENDED\_CLASS\_WAVG} & {= 3} \\
		\hline
		\multicolumn{1}{l}{FLAGS\_FOOTPRINT} & {= 1} \\
		\multicolumn{1}{l}{FLAGS\_FOREGROUND} & {= 0} \\
		\multicolumn{1}{l}{bitand(FLAGS\_GOLD,120)} & {= 0} \\
		\multicolumn{1}{l}{bitand(FLAGS\_BADREGIONS,1)} & {= 0} \\
		\hline
	\end{tabular}
	\caption{The flags used to select data in the DES Y3 GOLD catalogue. The first two flags guarantee that the astronomical objects which are the most likely to be a galaxy are selected, and the last four flags indicate the samples are clean, consistent with the Y3 GOLD footprint, and with} a reliable analysis from the {\sc SExtractor} \citep{Bertin1996}.
	\label{ch4_tab_flag}
\end{table}

\section{Convolutional Neural Networks (CNN)}
\label{ch4_sec_cnn}
\begin{figure*}{}
\begin{center}
\graphicspath{}
	\includegraphics[width=2\columnwidth]{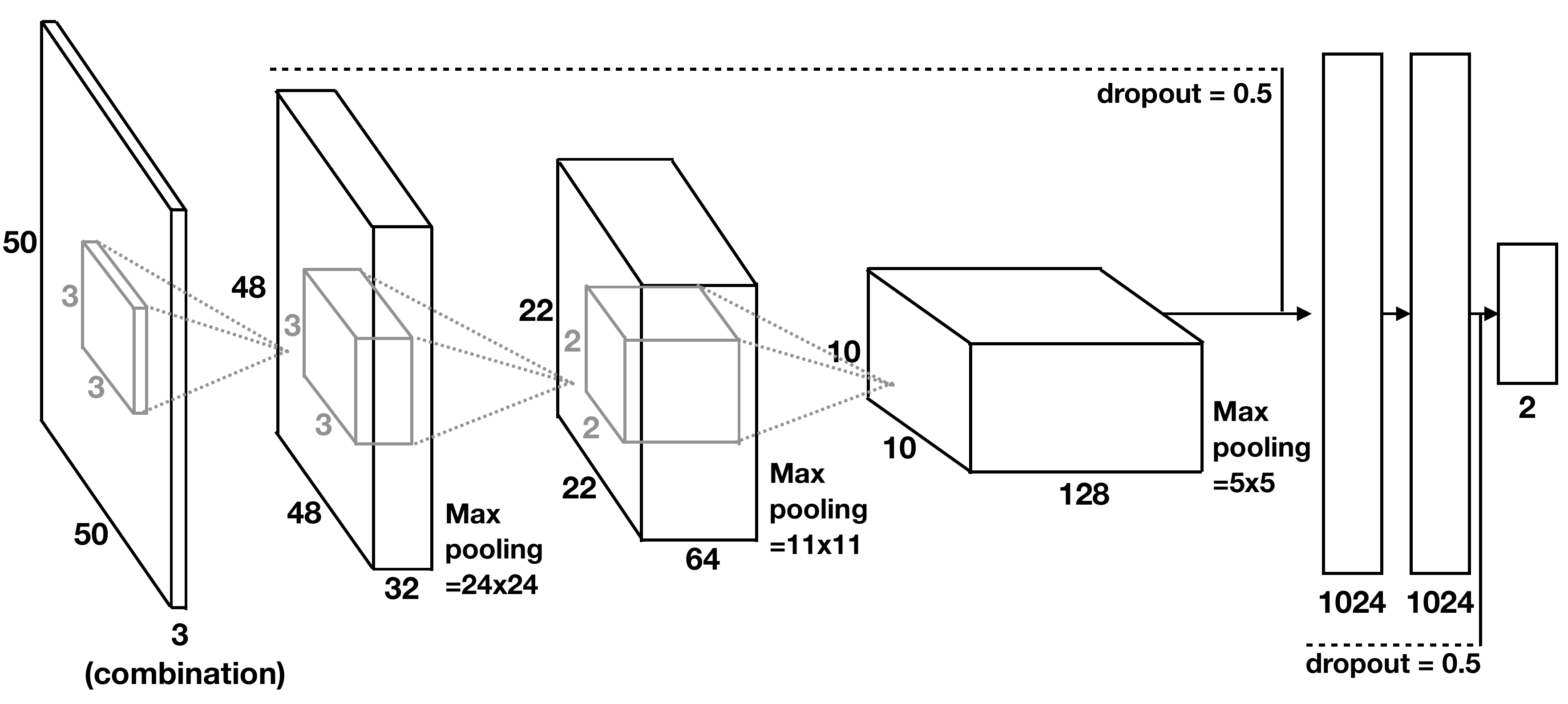}
   	\caption{The schematic overview of the CNN architecture used throughout. The architecture starts from an input of dimension $50\times50\times3$, and is followed by three convolutional layers with kernel sizes of 3, 3, 2 and channel sizes of 32, 64, 128, respectively plus a max-pooling layer after each. Two dense layers with 1,024 hidden units are following the third convolution layer. A dropout ($p=0.5$) is applied after the the third convolutional layer and after the second dense layer. Probabilities for two classes are predicted in the final output of our CNN, ‘Ellipticals’ and ‘Spirals’.}
    \label{ch4_fig_cnn}
\end{center}
\end{figure*}

Convolutional Neural Networks \citep[CNN, ][]{LeCun1998} are a type of neural network which includes convolutional layers used to extract strongly weighted features from input images for a given classification problem. The architecture of the CNN used throughout this paper is shown in Fig.~\ref{ch4_fig_cnn}. This design is inspired by the best performing architecture used in \citet{Dieleman2015}, but with fewer convolutional layers and parameters. The dimension of the inputs is $50\times50\times3$, with the depth including the linear images, HOG images, and log images. Three convolutional layers with kernel sizes of 3, 3, 2, respectively, are used in this study, and each of them is followed by a max-pooling layer with a size of 2. The max-pooling layer is also referred to as a `downsampling' layer, which is used to reduce the spatial size and the numbers of parameters involved in the architecture. After the third convolutional layer, two dense layers with 1,024 hidden units for each layer follow. In addition, dropouts ($=0.5$) are applied to reject irrelevant parameters and prevent overfitting in training the CNN. A dropout follows the third convolutional layer (max-pooling layer), and the other one comes after the two dense layers.

The activation function used in the convolutional layers and the dense layers is the Rectified Linear Unit \citep[{\texttt{ReLu}};][]{Nair2010_cs} such that $f(z)=0$ if $z<0$ while $f(z)=z$ if $z\ge0$. Finally, the {\texttt{softmax}} function \citep{Bishop2006}, $f\left( { z } \right) ={ \exp\left( z \right)  }/{ \sum { \exp\left( { z }^{ j } \right)  }  }$ is applied to the output layer and provides the probability distribution of each type.  For the CNN training, we apply {\it{Adam}} Optimiser, {\it{Nesterov momentum}}, and  set ${\it{momentum}}=0.9$ according to \citet{Dieleman2015}. The learning rate is set to 0.001, and the maximum number of iterations is 500, with an early-stopping mechanism that triggers when the validation set hits the local minimal loss.

A CNN has the technical advantage of not requiring the pre-processing procedure commonly used in artificial neural networks. However, in C20, we have proven that combining pre-processed images such as HOG images and log images qualitatively improves the performance of our CNN and reaches a final accuracy of over 0.99. Accuracy here is defined as the number of matched classifications by CNN and GZ1 from the total overlapped samples (Equation~\ref{ch4_eq:accuracy}). In this study, we independently train the CNN five times with the datasets described in Section~\ref{ch4_sec_traindata} which is then randomly separated into training and validation set with a fraction of 0.9 and 0.1 of the total, respectively in each run. Doing this avoids using exactly same batches for training each run. We then apply these pre-trained models to predict morphological classifications for the DES Y3 data (Section~\ref{ch4_sec_y3data}).  The final morphological prediction is obtained by averaging the predicted probabilities of these five independent CNN models for each type -- ‘Ellipticals’ and ‘Spirals’.

\section{Catalogues for Cross-validation}
\label{ch4_sec_cata4comp}

Once we have the morphological predictions from the convolutional neural network (CNN) for millions of galaxies, it is of great importance to validate the reliability of these classifications. In this study, we compare our CNN predictions with four different resources: (1) the Galaxy Zoo 1 (GZ1) catalogue using the galaxies that were not present in the training set (Section~\ref{ch4_sec_gz_cata}); (2) visual classifications carried out by TC, CC, and AAS\footnote{TC: Ting-Yun Cheng; CC: Christopher J. Conselice; AAS; Alfonso Arag\'on-Salamanca\label{shortname}} (Section~\ref{ch4_sec_visclf}); (3) VIPERS unsupervised spectral classification \citep[][Section~\ref{ch4_sec_vipers_cata}]{Siudek2018},  and (4) non-parametric methods using the structural measurements from \citet{Tarsitano2018} (Section~\ref{ch4_sec_des_y1cata}). In DES Y3 GOLD catalogue, a quantity with `FRACDEV' \citep{Everett2020}, which indicates the fraction of the fitted galaxy profile represented by a de Vaucouleurs model \citep{deVaucouleurs1948}, may be used to compare with our classification. However, there are some priors used in the assignment of this fraction which might need a further examination for its reliability. Therefore, in this work, we do not use this quantity to compare with our CNN classification. The validation between them could possibly be further investigated in the future work.



\subsection{Visual classification of randomly selected subsamples}
\label{ch4_sec_visclf}
Visual classification (hereafter, VIS) was carried out by three of the co-authors (TC, CC, and AAS\textsuperscript{\ref{shortname}}) for a reasonably large number of galaxies. We randomly selected 500 galaxies per magnitude bin from the DES Y3 dataset for galaxies with $16\le{i}\le22$. For the brightest bins ($16\le{i}<18$), only galaxies in GZ1 were included. In doing so, we covered the whole magnitude range of the DES sample with a significant overlap with GZ1 for cross-validation.

\begin{figure*}{}
\begin{center}
\graphicspath{}
	\includegraphics[width=2.1\columnwidth]{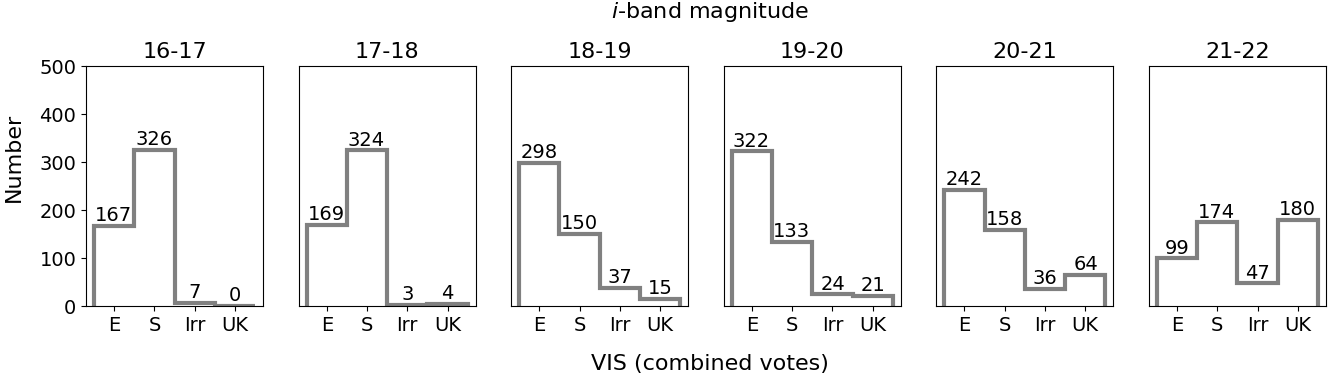}
   	\caption{The frequency distribution of each visual classification in each magnitude bin. On the x-axis, `E', `S', `Irr', and `UK' is short for Ellipticals, Spirals, Irregulars, and Unknown, respectively. The number above each bar represents the number of galaxies visually classified in that category. The range of the magnitude in $i$-band is shown at the top of each panel.}
    	\label{ch4_fig_vis}
\end{center}
\end{figure*}

The classification system we use is displayed in Table~\ref{ch4_tab_visclf}. We classify galaxies into six categories: Ellipticals (0), Early Spirals (1), Late Spirals (2), Edge-on Spirals (3), Irregulars (4), and Unknown (5). To compare with our CNN predictions, which provides probabilities for binary classification, for the Ellipticals and Spirals, we merge three subcategories of spiral galaxies into one - Spirals (1), and others retain the original label. The label with the most {\it combined votes} (Table~\ref{ch4_tab_visclf}) from our visual classifiers is set as the final visual type of a galaxy. This is the morphological type which is picked by at least 2 out of 3 of the classifiers. Those galaxies without a dominant label are categorised into the class of `Unknown'; the relative fraction of these `unknown' types increases with magnitude. The distribution of each visual type in each magnitude bin is shown in Fig.~\ref{ch4_fig_vis}.
\begin{table}
	\centering
	\begin{tabular}{lll}
		\hline
		\multicolumn{1}{l}{labels} & {primary votes} & {combined votes}\\
		\hline\hline
		\multicolumn{1}{l}{0} & {Ellipticals} & {Ellipticals} \\
		\multicolumn{1}{l}{1} & {Early Spirals} & {Spirals} \\
		\multicolumn{1}{l}{2} & {Late Spirals} & {} ''\\
		\multicolumn{1}{l}{3} & {Edge-on Spirals} & {} '' \\
		\multicolumn{1}{l}{4} & {Irregulars} & {Irregulars} \\
		\multicolumn{1}{l}{5} & {Unknown} & {Unknown} \\
		\hline
	\end{tabular}
	\caption{The classification system we applied in the visual classification carried out by TC, CC and AAS\textsuperscript{\ref{shortname}}. Galaxies are classified into 6 categories ({\it primary votes}) which are then merged into 4 categories, i.e. Ellipticals, Spirals, Irregulars, and Unknown ({\it combined votes}; see text).}
	\label{ch4_tab_visclf}
\end{table}

\begin{figure}{}
\begin{center}
\graphicspath{}
	\includegraphics[width=\columnwidth]{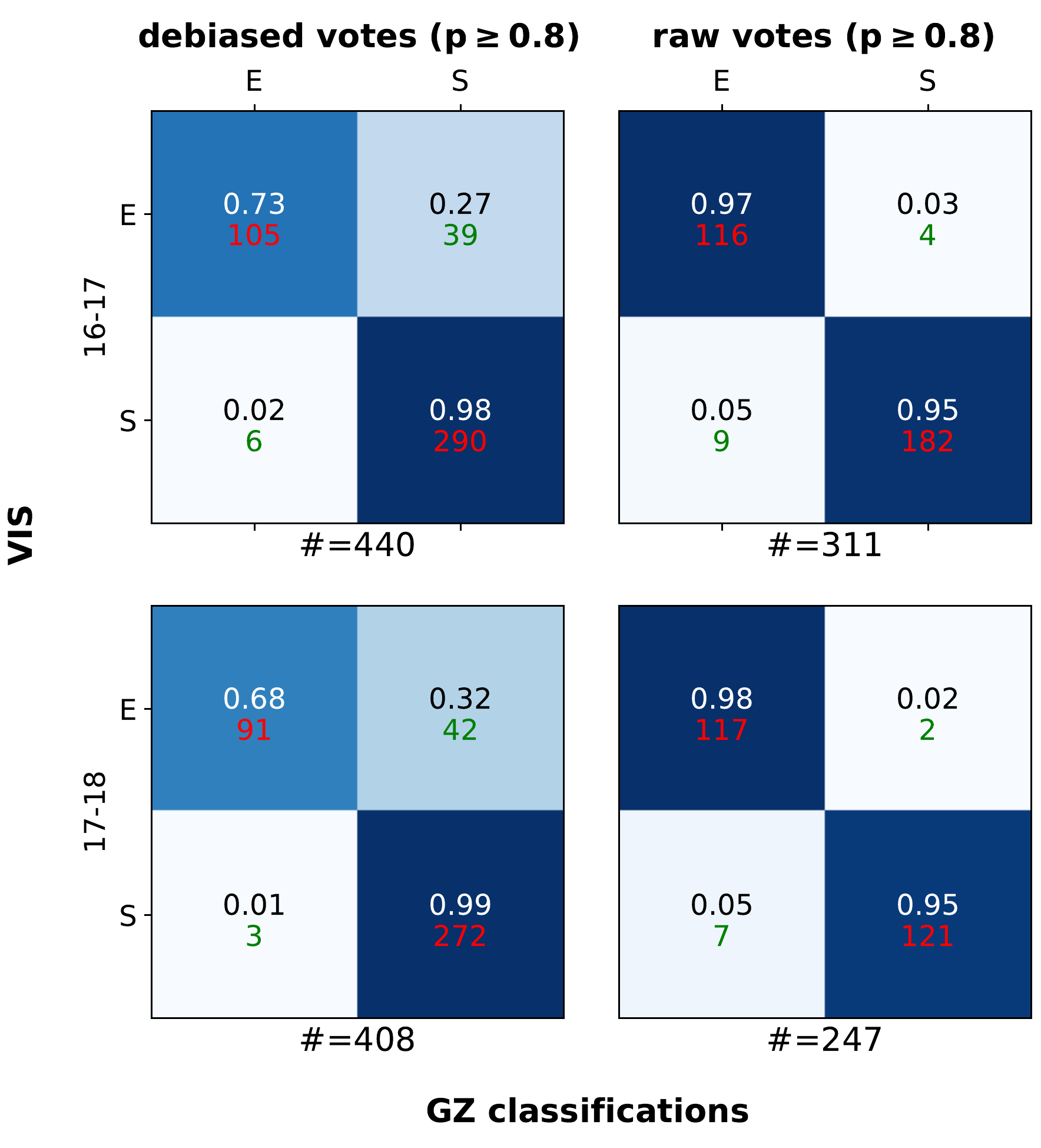}
   	\caption{The confusion matrices between our visual classifications (VIS) and the GZ1 classifications based on the {\it debiased votes} (first column) and {\it raw votes} (second column) \citep{Lintott2011}. A threshold of 0.8 is applied to both votes here to select high confidence classifications. Rows are separated by different magnitude bins: $16\le{i}<17$ (first row) and $17\le{i}<18$ (second row).}
    \label{ch4_fig_comp_GZ_vis}
\end{center}
\end{figure}

In order to validate the VIS, we compared the classifications of brighter galaxies ($i<18$) with the GZ1 classifications based upon the {\it debiased votes} and {\it raw votes} (Fig.~\ref{ch4_fig_comp_GZ_vis}). The {\it raw votes} directly reflect the votes from the volunteers of the GZ1. The {\it debiased votes}, as described in Section~\ref{ch4_sec_gz_cata}, are bias corrected using the E/S ratio measured directly from the raw likelihood. We apply a threshold of 0.8 to both votes to decide the morphology type with a higher confidence.

In Fig.~\ref{ch4_fig_comp_GZ_vis}, our VIS classifications show apparently better agreement with the raw votes from the GZ1 volunteers when comparing with the GZ1 debiased votes. The majority of the mismatched cases when comparing with the labels based on the debiased votes occur when a galaxy is classified as Elliptical by our visual classifications. This indicates that our judgement for galaxy morphology is also biased by the size, magnitude, and redshift of the galaxies. This gets worse when a galaxy is fainter which is shown in Fig.~\ref{ch4_fig_vis}. It is clear that significantly more galaxies are visually classified as Ellipticals.

Although our visual classification suffers from the same type of biases as GZ1, unfortunately we cannot perform a bias correction similar to the one they carried out. There are several reasons for this. First, the broader redshift range of our sample makes the assumption of unevolving morphological mix unreliable. Second, the number of galaxies we have been able to classify is too small to provide reliable correction statistics. And third, the lack of spectroscopic redshifts would render any redshift-dependent correction highly uncertain. Therefore, additional factors such as S\'ersic index \citep{sersic1963, sersic1968} and colour will be considered to validate the CNN predictions (Section~\ref{ch4_sec_compvis}).

\subsection{Unsupervised spectral classification}
\label{ch4_sec_vipers_cata}
With the known correlation between spectral classification of galaxies and galaxy morphology \citep[][etc]{Morgan1957,Bershady1995,Zaritsky1995,Kennicutt1998,Baldry2004}, we compare our predictions within a fainter magnitude ($i\ge18$) with an unsupervised spectral classifications presented in \citet{Siudek2018} from the VIMOS Public Extragalactic Redshift Survey (VIPERS). This provides a different way to examine the robustness of our CNN predictions, although with some caveats. These spectral classifications employ a Fisher Expectation-Maximization (FEM) unsupervised algorithms to categorise galaxies with redshift $z\sim0.4-1.3$ into 12 classes based on 12 rest-frame magnitude and spectroscopic redshift. Except for the class 12, which are the class of broad-line active galactic nuclei, other classes can be classified into three main categories: passive (class 1-3), intermediate (class 4-6), and star-forming galaxies (class 7-11).

\subsection{DES Y1 catalogue of morphological measurements}
\label{ch4_sec_des_y1cata}
To obtain a reliable analysis of the quality of our CNN labels, parametric factors such as the S\'ersic index and non-parametric coefficients such as {\it CAS system} (Concentration, Asymmetry, and Smoothness/Clumpiness), Gini coefficient, and M20 are used in this study. \citet{Tarsitano2018} included 45 million objects selected from the first year DES data, and provided the largest structural catalogue to date for galaxies. The selected samples from this catalog cover the magnitude range of $i\le23$.  According to the suggestions from the paper, we apply an initial cut as follows:
\begin{itemize}
    \item MAG\_AUTO\_I $\le$ 21.5
	\item SN\_I $>$ 30
	\item SG $>$ 0.005,
\end{itemize}
where {\it MAG\_AUTO\_I} represents the cut in $i$-band apparent magnitude and $\it SN\_I$ is the signal-to-noise ratio in the $i$-band. The $SG$ is used for optimising the separation between stars and galaxies while maintaining the completeness. The cut ({\it SG} > 0.005) recommended in \citet{Tarsitano2018} is the optimal compromise between the completeness and purity of the galaxy sample. These selections provides 12 million galaxies with 90\% completeness in S\'ersic measurements and 99\% completeness in non-parametric measurements in the $i$-band.

The parameters provided from the single S\'ersic fits (e.g. S\'ersic index, ellipticity, etc) are measured with {\sc Galfit} \citep{Peng2010}. We then apply a further cut suggested in \citet{Tarsitano2018} to select the galaxies that are successfully validated and calibrated. The calibration is made based upon four parameters: size, magnitude, S\'ersic index, and ellipticity using simulated galaxies generated with these parameters \citep{Tarsitano2018}:
\begin{itemize}
	\item FIT\_STATUS\_I = 1
\end{itemize}
On the other hand, the non-parametric parameters (CAS parameters, Gini, and M20) are measured using the Zurich Estimator of Sturctural Types ({\sc ZEST+}) \citep{Scarlata2007a, Scarlata2007b}. The calibration is applied with the same procedure as the parameter fit but uses concentration instead of S\'ersic index for non-parameteric parameters, and the validation is discussed on the Gini-M20 plane as a function of other morphological measurement such as concentration (C), asymmetry (A), and clumpiness (S) \citep{Tarsitano2018}. One criterion is applied in non-parametric coefficients to select the objects with successfully validated and calibrated measurements.
\begin{itemize}
	\item FIT\_STATUS\_NP\_I = 1
\end{itemize}

\section{Validation \& Discussion}
\label{ch4_sec_validation}

In this section, we carry out the cross-validation of our CNN predictions using multiple sources listed in Section~\ref{ch4_sec_cata4comp}. Included among this, we also discuss the confidence levels assigned to the predictions with a probability threshold of 0.5 and the uses of this catalogue with these confidence levels. That is, we explain how to use our catalog for determining galaxy morphologies.

Some quantities are used to examine the performance of our CNN classifications such as accuracy, precision (Prec), recall (R), true positive rate (TPR; the same definition as R), and false positive rate (FPR). Accuracy is defined as the number of correct classifications compared to the `true' labels from the total samples. In Equation~\ref{ch4_eq:accuracy}, `T' and `F' represents `True' and `False' while `P' and `N' denotes `Positive' and `Negative', respectively.
\begin{equation}
\label{ch4_eq:accuracy}
    	Accuracy=\frac { TP+TN }{ TP+FP+TN+FN }.
\end{equation}
Precision and recall are defined as follows,
\begin{equation}
\label{ch4_eq:prec_rec}
    	R=\frac { TP }{ TP+FN } ;\quad Prec=\frac { TP }{ TP+FP }.
\end{equation}
Finally, true positive rate (same definition as R) and the false positive rate are defined as below,
\begin{equation}
\label{ch4_eq:cm_tp_fn}
    	TPR=\frac { TP }{ TP+FN };\quad FPR=\frac { FP }{ FP+TN } .
\end{equation}

\subsection{Galaxy Zoo 1 catalogue (GZ1)}
\label{ch4_sec_compgz}

\begin{figure}{}
\begin{center}
\graphicspath{}
	\includegraphics[width=\columnwidth]{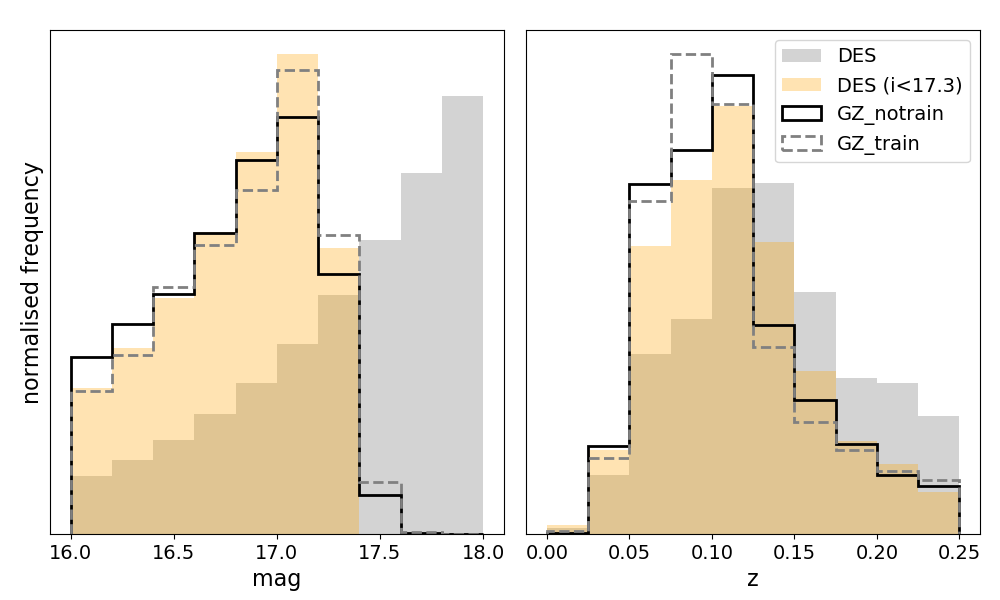}
   	\caption{The magnitude and redshift distribution of the DES Y3 data with the same coverage as the training set (Section~\ref{ch4_sec_traindata}). The gray and yellow shading represents the DES Y3 data without and with a cut at $i=17.3$, respectively. The solid lines show the overlap region with the GZ1 catalogue, excluding the training set, while the dashed lines show only the training set.}
    \label{ch4_fig_gz_cnn_data_dist}
\end{center}
\end{figure}

To validate our CNN predictions, first we compare our CNN classifications with the GZ1 labels based upon the {\it debiased votes} (Section~\ref{ch4_sec_gz_cata}). We do this by matching our DES Y3 data with the GZ1 catalogue which provides us with $\sim$2,700 additional galaxies that are not within the training set. These additional samples are used to examine our CNN predictions in this section. The distribution of the DES Y3 data for this test is in the same magnitude and redshift range as the training set (Section~\ref{ch4_sec_traindata}) as shown in Fig.~\ref{ch4_fig_gz_cnn_data_dist}. Note that there are significantly fewer faint galaxies at $i>17.3$ in our sample with overlapping GZ1 classifications. Therefore, a cut of $i=17.3$ is applied when carrying out the analysis in this subsection. We then discuss the performance of the CNN predictions below and above this magnitude limit in later sections.

\begin{figure*}{}
\begin{center}
\graphicspath{}
	\includegraphics[width=2\columnwidth]{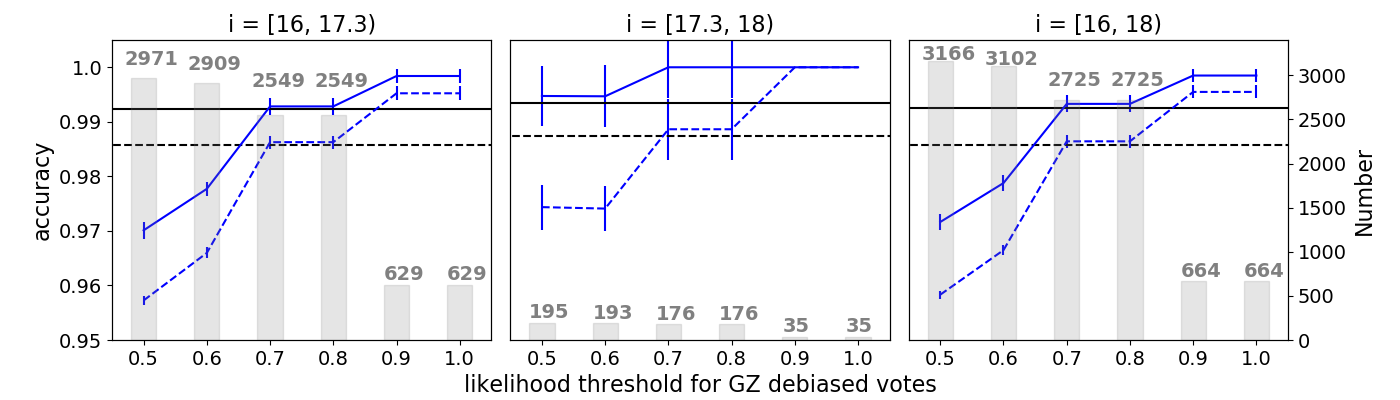}
   	\caption{The comparison between the accuracy of the CNN predictions when applying different likelihood thresholds to the GZ1 debiased votes. A magnitude cut $i=17.3$ is applied in the first two panels, and the third panel includes total data in the first two panels. The black lines are the accuracy of the training sets when applying a probability threshold, $p=0.5$ (dashed line) and $p=0.8$ (solid line) to our CNN predictions.  The blue lines represent the change of accuracy when comparing our CNN predictions, based on a probability cut $p=0.5$ (dashed line) or $p=0.8$ (solid line), with the GZ1 classifications based on different likelihood thresholds. The second y-axis on the right reflects the height of the shading bars which gives the number of data points left after applying a likelihood threshold. The error bars represent the standard deviation of the accuracy obtained within five models.}
    \label{ch4_fig_accuracy_gz_threshold}
\end{center}
\end{figure*}

First, in Fig.~\ref{ch4_fig_accuracy_gz_threshold}, we show the change in accuracy when applying different likelihood thresholds to the GZ1 debiased votes. The first two panels are separated by the magnitude cut $i=17.3$, and the third panel contains all overlapping data between GZ1 and DES Y3 data used in this paper. The accuracy of the training set is represented by the black lines. One applies a probability threshold of $p=0.5$ to our CNN predictions (dashed line), the other applies a threshold of $p=0.8$ (solid line). The comparison of the results using different likelihood threshold at various GZ1 debiased votes are shown by the blue lines. The line styles reflect the same meaning as the black lines. Meanwhile, the second y-axis is used for the shading bars which show the number of galaxies under the likelihood threshold. Both Prec and R, as defined in Equation~\ref{ch4_eq:prec_rec}, of each datapoint in Fig.~\ref{ch4_fig_accuracy_gz_threshold} are very high. For example, both Prec and R are $\ge0.97$ at each data point when CNN uses $p=0.5$, while the two values are $\ge0.98$ with $p=0.8$ on CNN.

We note that the accuracy of our CNN predictions compared with the GZ1 classifications based upon a debiased likelihood threshold of 0.8 shows a good consistency with the accuracy of the training set (the first panel in Fig.~\ref{ch4_fig_accuracy_gz_threshold}). In the second panel ($17.3\le{i}<18$), the CNN shows a slightly better performance than the brighter range ($16\le{i}<17.3$) and training set. However, the scatter for the CNN predictions are larger because there are significantly fewer samples in the second plot. When taking the scatter into account, the performance of our CNN predictions in this magnitude range also shows a good consistency with the training set. Therefore, based upon Fig.~\ref{ch4_fig_accuracy_gz_threshold}, we interpret that there is a `superior confidence' level to the CNN predictions within the brighter magnitude range  $16\le{i}<18$ and redshift range $z<0.25$. Additionally, in later analysis, we apply a likelihood threshold of 0.8 to the GZ1 debiased votes to determine the GZ1 classifications for comparison. In our catalogue, we provide a  classification flag with a probability threshold of 0.8 (89\% of the total samples; {\it MORPH\_FLAG} in Table~\ref{ch4_tab_catalog_col}) to reject samples with low predicted probabilities from the CNN model. The prediction with a probability lower than 0.8 is labelled as `uncertain (-1)' in our catalogue (Section~\ref{ch4_sec_y3catalog}). Several reasons might result in the low predicted probabilities which has been discussed in C20. One of the possible reasons is caused by stellar contamination in the DES galaxy sample at $i<18$.

\begin{figure*}{}
\begin{center}
\graphicspath{}
	\includegraphics[width=2\columnwidth]{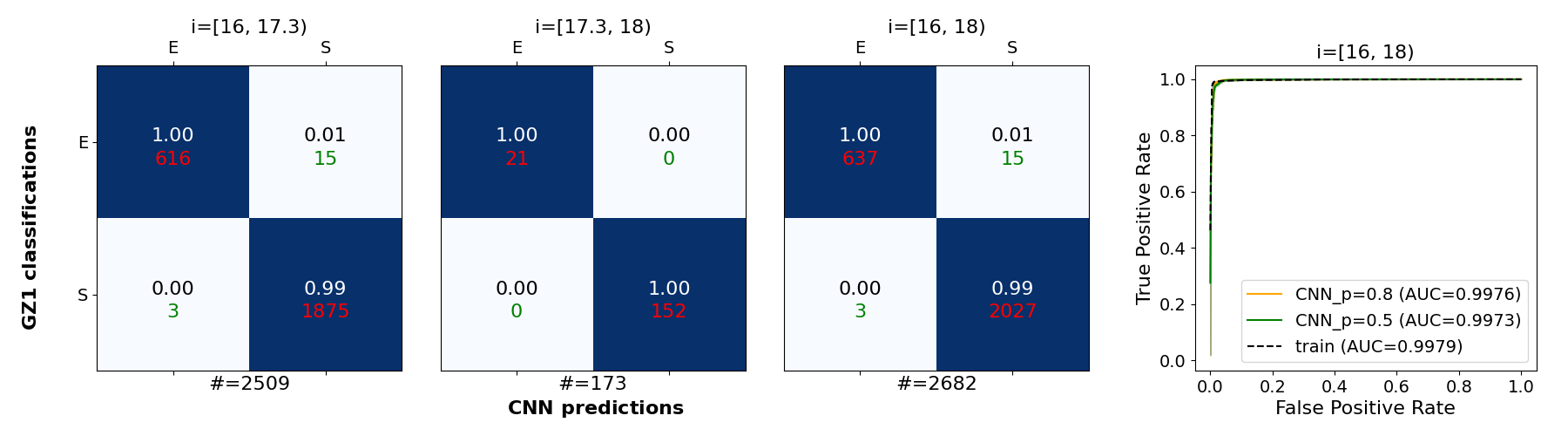}
   	\caption{The combined graph of the confusion matrices and the ROC curve comparing our CNN predictions with the GZ1 labels defined by the debiased votes with a threshold of 0.8. The first three panels show the confusion matrices within certain magnitude ranges: $16\le{i}<17.3$, $17.3\le{i}<18$, and $16\le{i}<18$, where the CNN predictions use a probability threshold of 0.8. The red or green colour in each quadrant represent the number of galaxies that agree with the classifications derived through CNN predictions and GZ1 classifications within each quadrant. The number above it indicates the fraction of these galaxies within each certain type decided by our CNN classifier. The number of galaxies within each magnitude range is shown below each graph.} The last panel shows the ROC curve where the {\it y}-axis represents the true positive rate, and the {\it x}-axis is the false positive rate. The orange and green lines show the curve from our CNN predictions with  probability thresholds of 0.8 and 0.5, respectively, while the black dashed line is from the training set.
    \label{ch4_fig_roc_cm_GZ_cnn}
\end{center}
\end{figure*}

In the first three panels of Fig.~\ref{ch4_fig_roc_cm_GZ_cnn}, we show confusion matrices within a certain magnitude range as listed above the graph. The {\it x}-axis indicates the CNN predictions with a probability threshold of 0.8, while the {\it y}-axis shows the GZ1 classifications with a vote threshold of 0.8. We focus on classifying galaxies into two types, Ellipticals (E) and Spirals (S). The numbers at the bottom of the confusion matrices show the number of galaxies that are not classified as uncertain type within the ranges in each column. For the first three plots, we exclude the training set, and compare the performance with the training set in the last panel. In this figure, we notice that the two labels (GZ1 and CNN) match well, and the majority of mismatches occur in the case where the CNN classification is Spiral, but the debiased GZ1 classification disagrees.

\begin{figure}{}
\begin{center}
\graphicspath{}
	\includegraphics[width=\columnwidth]{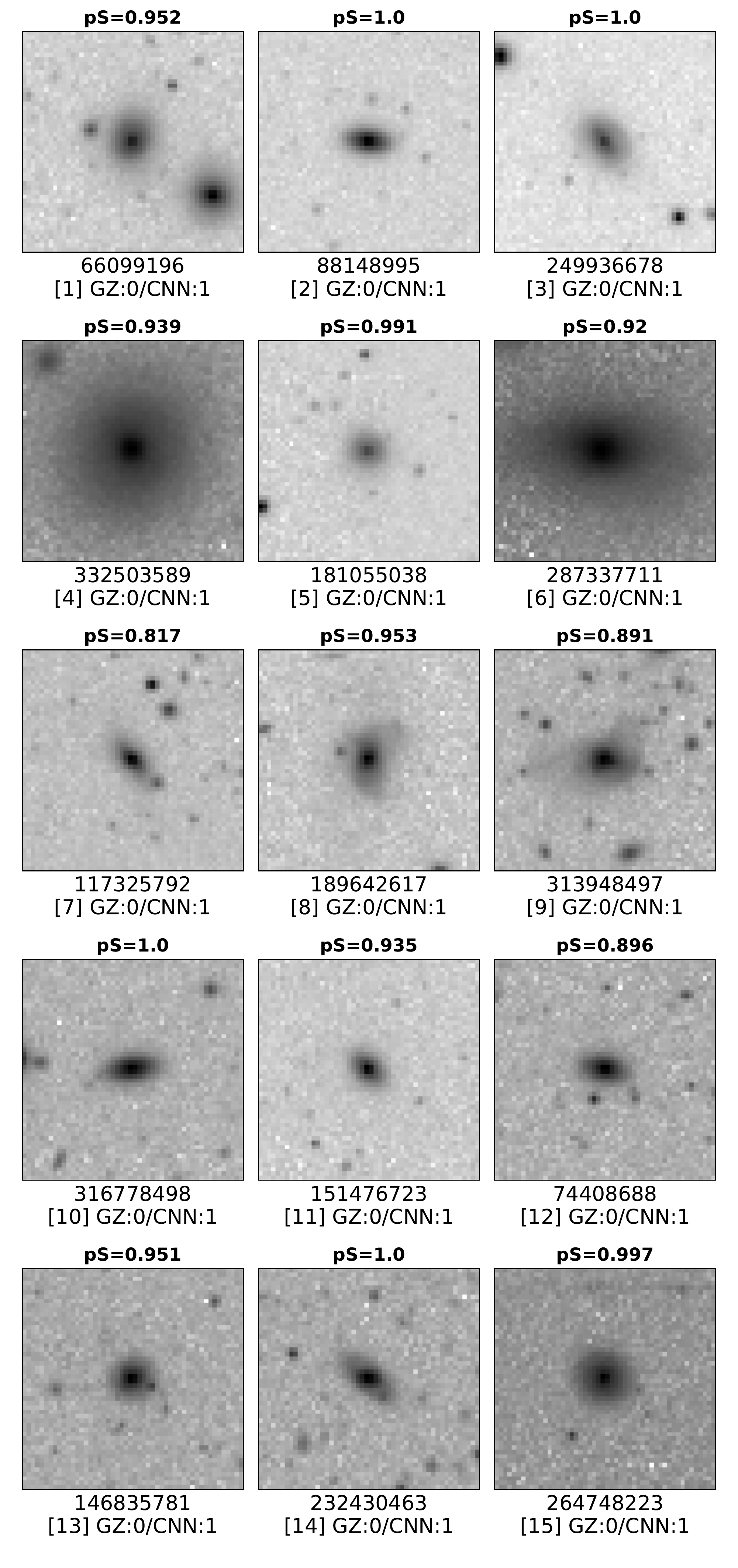}
   	\caption{Examples of galaxies that our CNN classified as Spirals while the GZ1 labelled as Ellipticals. The predicted probability of being Spirals from the CNN is shown above each stamp ({\bf pS}).}
    \label{ch4_fig_example_mismatch_GZ_cnn}
\end{center}
\end{figure}

Fig.~\ref{ch4_fig_example_mismatch_GZ_cnn} showcases the galaxies which are classified as Spirals by the CNN but Ellipticals by the GZ1. Some galaxies in this category show disky structures (e.g. [1], [3], [5], [14], [15]) or asymmetric features (e.g. [8] and [9]) in the DES imaging data. In C20, we proved that the higher quality DES imaging data reveals detailed structures that were not detected in the data from SDSS. This condition explains the mismatched classifications happened in Fig.~\ref{ch4_fig_example_mismatch_GZ_cnn}.

Ideally, we would have liked to examine the S\'ersic index distribution of these mismatched galaxies to determine whether these misclassified galaxies have particular structural properties. However, in this case, there are fewer than three overlapping galaxies with mismatched labels in \citet{Tarsitano2018}. The mismatched test sample is far too small for any statistically meaningful analysis. Therefore, we leave this additional cross-validation to future work, when more structural measurements are obtained for the DES Y3 data. We note that, although we cannot carry out this additional test, we are confident of the excellent performance of our CNN predictions within the magnitude and redshift range covered by the GZ1 training set.  Based on the discussions above and, in particular, the confusion matrices shown in Fig.~\ref{ch4_fig_roc_cm_GZ_cnn}, we conclude that in this magnitude and redshift range, which includes $\sim670,000$ galaxies, our CNN classifier has an accuracy of over 99\%.

The last panel in Fig.~\ref{ch4_fig_roc_cm_GZ_cnn} shows a Receiver Operating Characteristic curve \citep[ROC curve,][]{Fawcett2006, Powers2011} which is used to examine the performance of our machine learning technique by comparing the probabilities predicted by the machine with the true labels. On a ROC curve the \textit{y}-axis is the TPR and the \textit{x}-axis is the FPR (Equation~\ref{ch4_eq:cm_tp_fn}); therefore, the closer a ROC curve gets to the corner (0,1), the better the performance is.

Another important indicator on the ROC curve is the `area under the curve', AUC, which has a larger value for a better performance of a machine learning model. From the ROC curve, both CNN predictions with a probability thresholds of 0.5 (green line) and 0.8 (orange line) within the coverage of the training sets in magnitude ($16\le{i}<18$) and redshift ($z<0.25$) show an excellent consistency with the results of the training set. This result doubly confirms our confidence on these predictions. Therefore, the CNN predictions within this range are labelled as `superior confidence (4)' in the {\it confidence\_flag} (Table~\ref{ch4_tab_confidence_flag}) in the catalogue, Table~\ref{ch4_tab_catalog_col}.

\subsection{Visual classification}
\label{ch4_sec_compvis}

\begin{figure*}{}
\begin{center}
\graphicspath{}
	\includegraphics[width=2.1\columnwidth]{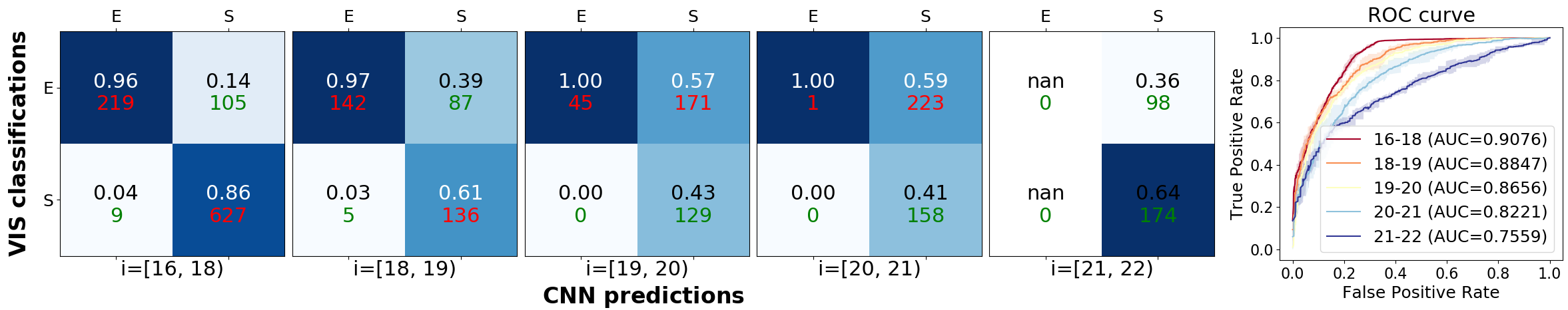}
   	\caption{The confusion matrices and the ROC curve of different magnitude ranges. The {\it x}-axis of the confusion matrices is the CNN predictions with a probability threshold of 0.8 and the {\it y}-axis is our visual classifications. On the ROC curve, the {\it x}-axis is the false positive rate while the {\it y}-axis represents the true positive rate. Different colours indicate different magnitude ranges.}
    \label{ch4_fig_roc_cm_vis_cnn}
\end{center}
\end{figure*}

To allow us to test the quality of the CNN classifications of fainter galaxies, we carried out a visual classification of 500 randomly picked galaxies in each $i$-band apparent magnitude bin with an interval of 1 magnitude using the DES imaging data. The first five panels in Fig.~\ref{ch4_fig_roc_cm_vis_cnn} show the confusion matrices in each magnitude range, and the ROC curve is shown on the last panel. We apply a probability threshold of 0.8 to our CNN predictions in this figure to reject samples with low predicted probabilities. The performance quality of our CNN method drops with magnitude when comparing with the visual classifications.

\begin{figure*}{}
\begin{center}
\graphicspath{}
	\includegraphics[width=2.1\columnwidth]{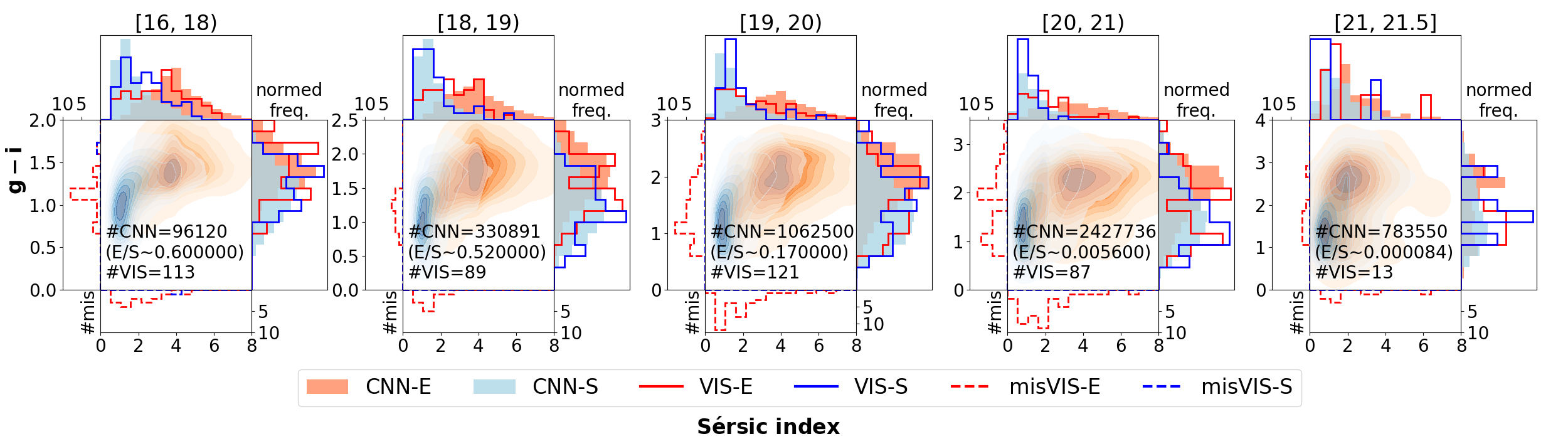}
   	\caption{The diagram shows the colour and the S\'ersic index distribution of our samples. The {\it y}-axis presents the colour $g-i$ and the {\it x}-axis shows the S\'ersic index. The central contour plot shows the two dimensional density distribution while the histograms at the top and the right show the normalised frequency distribution of each quantity, S\'ersic index and $g-i$ colour, respectively. The histograms at the bottom and the left show the samples with mismatched labels between the CNN and visual classifications for the S\'ersic index and the colour, respectively. The red/orange colour represents  the Ellipticals, while the bluish colour is for the spiral galaxies. The shadings represents the CNN predictions, and the solid lines show the distribution labelled by the visual classifications. Finally, the dashed lines show the misclassified samples with the labels from the visual classifications.}
    \label{ch4_fig_color-sersci_vis_cnn}
\end{center}
\end{figure*}

There are two main reasons responsible for this decreasing performance. First of all, through the confusion matrices, we notice that the majority of mismatches happened in the cases where our CNN method classified a galaxy as a spiral galaxy but we visually classified it as an elliptical galaxy. This situation is caused by the fact that our CNN is trained with the corrected debiased GZ1 classifications (Section~\ref{ch4_sec_gz_cata}); however, the visual classification used here is a raw classification. In Section~\ref{ch4_sec_visclf}, we pointed out that our visual classifications suffer from a similar classification bias compared with the raw GZ1 classifications which are influenced by the magnitude, size, and redshift of the targets. Therefore, in Fig.~\ref{ch4_fig_color-sersci_vis_cnn}, we combine the S\'ersic index and colour of each galaxy to cross-validate our results. The colour information is obtained using apparent magnitude, which is measured in an elliptical aperture determined by the Kron radius, from the DES Y3 GOLD catalogue. In this work, we use apparent colour for our validation instead of the colour with absolute magnitude. It is due to the large uncertainties in redshift estimation for the DES galaxies which can have a strong effect on absolute magnitude derivation. The S\'ersic index is from the DES Y1 morphological measurements \citep{Tarsitano2018} selected based upon the suggested flags  (described in Section~\ref{ch4_sec_des_y1cata}). Due to the applied cut in magnitude up to $i=21.5$ used in \citet{Tarsitano2018}, the last panel in Fig.~\ref{ch4_fig_color-sersci_vis_cnn} only shows galaxies within the magnitude range of $21\le{i}\le21.5$.

In Fig.~\ref{ch4_fig_color-sersci_vis_cnn}, the central contour shows the density distribution of the S\'ersic index and the ($g-i$) colour at each magnitude. The histograms at the top and the right show their respective normalised frequency distribution. The bottom and left histograms show the misclassified samples colour-labelled by the visual classifications. From this figure, it is clear that the majority of misclassified galaxies labelled as Ellipticals by our visual assessment are in fact diskier and bluer. Since our CNN is self-debiased by training with the corrected debiased GZ1 labels (Section~\ref{ch4_sec_gz_cata}), it shows a more sensible classification of the images than humans have difficulty to classify correctly. That is, our CNN classifications are more likely to be correct than the visually based ones.

We remind the reader that our CNN classifier is trained with monochromatic $i$-band images, without any colour information. Therefore, the strong colour segregation between CNN-classified Ellipticals and Spirals is reassuring: the connection between CNN morphology and colour is independent, and not based on the training process -- colour and galaxy morphology are linked through galaxy formation and evolution processes, and are not strongly the result of classification biases.

Second, in addition to the bias in the visual classification, another potential reason for the decreasing performance in our CNN classifier is caused by the fact that we train our CNN with brighter galaxies ($16\le{i}<18$) which are at a lower redshift ($z<0.25$), and then use this model to predict galaxies in different domains (i.e., galaxies with a fainter magnitude and a higher redshift). Combining with the `self bias correction' feature shown in our CNN, we expect that our CNN classifies more disk galaxies at fainter magnitudes. For example, for the faintest magnitude range in our study ($i\ge21$), the `self bias correction' of our CNN classifier is over applied due to the very low signal-to-noise ratio compared with the training set. This overdone bias correction gives us an artificially low number of Ellipticals classified by the CNN. The ratio of the CNN-classified Ellipticals to Spirals in this magnitude range is $\sim6\times{10^{-5}}$ for total samples and $\sim8.4\times{10^{-5}}$ for overlapped samples shown in Fig.~\ref{ch4_fig_color-sersci_vis_cnn}. The evolution of the E/S ratio strongly depends on the methods used to classify galaxy morphology, in particular at a high redshift. However, there is not a significant evolution in morphology mix within the redshift range in our sample (over $99.9\%$ of the galaxies at $z\le1.2$; Section~\ref{ch4_sec_y3data}). shown in previous studies \citep[e.g.,][]{Conselice2005, Cassata2005}. Therefore, the significant difference in the number of our CNN-classified Ellipticals and Spirals is likely caused by the reason discussed above.

This is shown in both the confusion matrix and the colour-S\'ersic diagram: no visually classifiable Ellipticals is picked out by our CNN classifier (Fig.~\ref{ch4_fig_roc_cm_vis_cnn}), and there is not a clear separation between Ellipticals and Spirals in the S\'ersic index distribution (Fig.~\ref{ch4_fig_color-sersci_vis_cnn}).

Machine learning is sensitive to image qualities such as the signal-to-noise ratios and resolution. In our case, the apparent magnitude of a galaxy, which is influenced by the redshift, affects the signal-to-noise ratio of the galaxy, which can affect how easily structure can be seen. Additionally, due to the effects of distance, a galaxy at a higher redshift shows less detailed structure, namely the resolution of the galaxy images decreases. However, there is a certain level of tolerance for variations within these effects, which is still a popular topic to investigate in computational science using  images for topics such as object identification, face recognition, etc \citep[][etc]{Dodge2016,Karahan2016,Amirshahi2016,Zhou2017,Prakash2019}.

For galaxy morphology, a few specific features such as light distribution, spiral arms, disk structures, etc, are dominant when visually classifying galaxies. This gives the possibility of using visual classification in galaxy morphology with images of a low quality. Similar to visual classification, our CNN shows a capability to classify galaxies based on the feature of light distribution and disk structure in Fig.~\ref{ch4_fig_color-sersci_vis_cnn}, which even shows a likely better classification than human opinion.

Additionally, we note that using monochromatic images we are sampling different rest-frame morphologies at different redshifts. An $i$-band image for an object at a higher redshift, e.g., $z=1$ (the upper limit in our final catalogue),
examines the morphology at $\sim$400 nm. One can debate if this fact helps or obscures machines in classifying galaxy morphologies at a higher redshifts. The different distributions presented in different rest-frame morphologies due to the redshift effect challenge the machine to adapt the domain learned in the training set to a different domain. However, bluer rest-frame morphologies emphasise the feature of spiral arms (location of young stars or star forming regions), which is one of the dominant feature in separating Ellipticals and Spirals. This fact might help the machine to distinguish Spirals even though the image resolution drops at a higher redshift.

Therefore, we statistically investigate the confidence of our CNN predictions at fainter magnitudes and higher redshifts by comparing the quality of our morphologies with our visual assessments, structural measurements such as the S\'ersic profile, and galaxy properties such as colour. This analysis also investigates the limit of our CNN classifier, which is trained with bright galaxies at low redshift, on classifying galaxies within different ranges of magnitude and redshift. The detailed discussion of this is in Section~\ref{ch4_sec_confidence_level}.

As a side note,  within the faintest magnitude bin in Fig.~\ref{ch4_fig_color-sersci_vis_cnn}, even though the CNN-classified Ellipticals are rare and do not have the expected S\'ersic index distribution, we still find a fairly good separation in their colour distribution. This indicates that the CNN-classified Ellipticals with $21\le{i}\le21.5$ share some similarities among themselves. Therefore, this particular class of galaxies might have a different formation history from other Ellipticals, resulting in a relatively disky structure but redder colours. It would be interesting to test this hypothesis with multicolour data in the future.

Nonetheless, based on the analysis in this section, we exclude the CNN classifications in the magnitude range ($i\ge21$) from our final catalogue due to the strong imbalance of the CNN classifications between the two types and the poor division in the colour-S\'ersic diagram.


\subsection{Further investigation to fainter galaxies}
\label{ch4_sec_confidence_level}
With the predicted probabilities provided by our CNN classifier, users can simply use these labels to allocate a classification with a higher predicted probability to a galaxy. However, as discussed above, the machine is trained with bright galaxies ($16\le{i}<18$) at low redshift ($z<0.25$), one might thus consider that the predictions for galaxies with similar properties as the training set are more robust. Therefore, in this section, we provide two additional quantities that can be used to select the CNN classifications that users might have more faith in based on different presumptions. We carry out the examination by further exploring our CNN predictions using S\'ersic index and colour ($g-i$). In this section, we statistically assess our CNN classifications with a probability threshold of 0.5 for each magnitude and redshift range.

One way to determine the quality of our classifications is based on using only the S\'ersic index, which is a fairly good indicator for galaxy structure even at high redshift. The predictions might be more robust if a CNN-classified Ellipticals has a S\'ersic index larger than 2.5 or a CNN-classified Spirals has S\'ersic index smaller than or equivalent to 2.5. Therefore, we statistically examine how well the CNN classifications do within a certain magnitude and redshift range to satisfy the corresponding S\'ersic index as discussed above. Equation~\ref{ch4_eq:sersic_confidence} is then used to provide `frac\_n' in Table~\ref{ch4_tab_catalog_col}, where N represents the total number of samples with S\'ersic index measurements and n means S\'ersic index. The pE and pS are the predicted probabilities of being Ellipticals and Spirals by CNN, respectively.
\begin{multline}
    \label{ch4_eq:sersic_confidence}
    frac\_n =\\
    \frac{N\left [ \left (pE>0.5 \right) \wedge \left (n>2.5\right) \right ]+N\left [ \left (pS>0.5 \right) \wedge \left (n\leq2.5\right) \right ]}{N}
\end{multline}

\begin{figure*}{}
\begin{center}
\graphicspath{}
	\includegraphics[width=2.1\columnwidth]{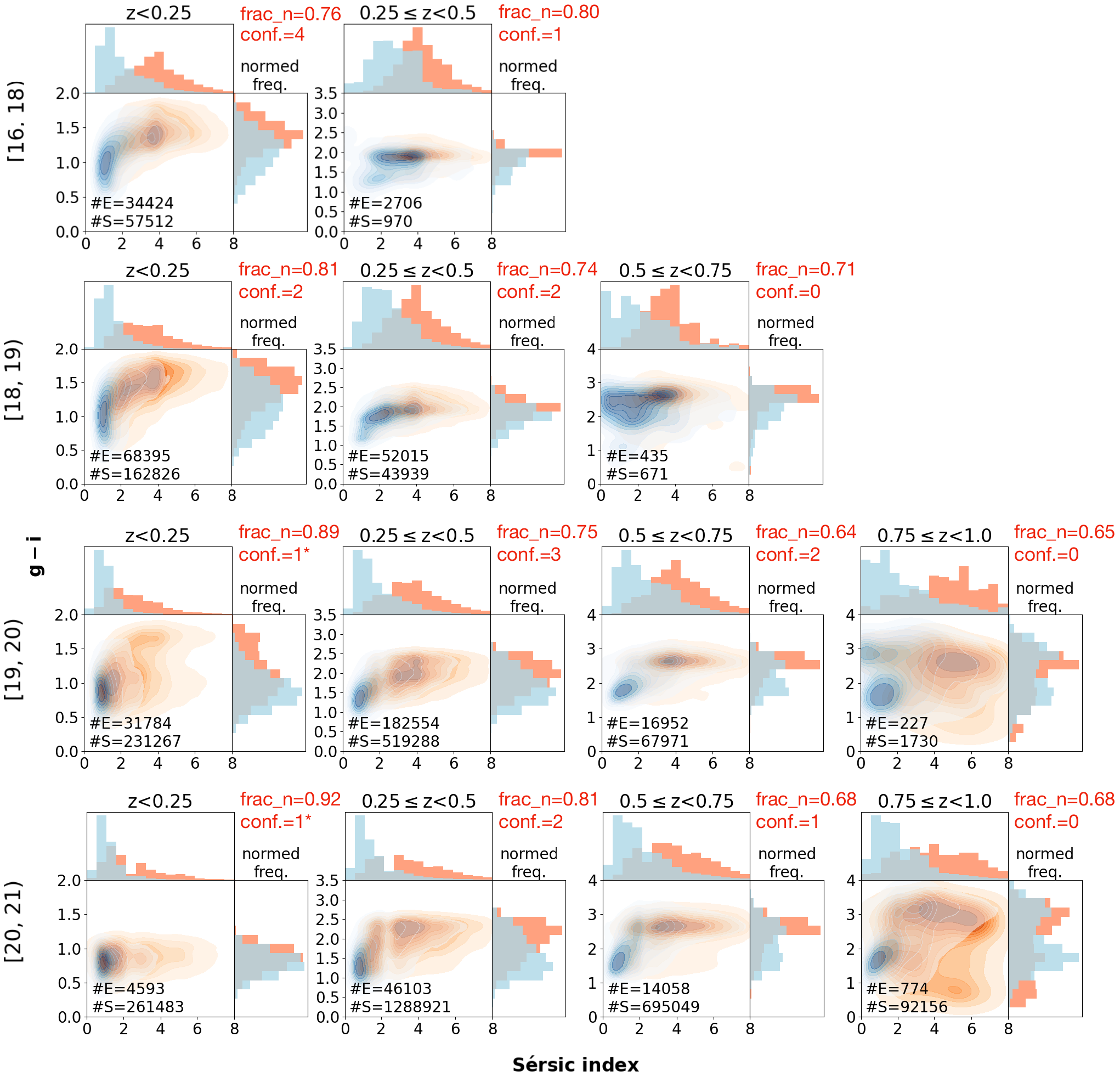}
   	\caption{The colour-S\'ersic diagrams of different redshift bins for each magnitude ranges. The histograms at the top and the right of each diagram show the normalised frequency distribution of S\'ersic index and colour $g-i$, respectively. The red shading represents the Ellipticals classified by our CNN with a probability threshold of 0.5, while the blue shading shows the CNN-classified Spirals with a probability threshold of 0.5. The magnitude range is shown at the left of each row while the redshift range is presented above each graph. The textual information in the diagrams shows the number of Ellipticals (E) and Spirals (S) classified by our CNN and with the DES Y1 morphological measurements from \citet{Tarsitano2018}. The red text at the top right corner of each plot indicates (1) the fraction of CNN classifications which satisfies S\'ersic index criteria (frac\_n; Equation~\ref{ch4_eq:sersic_confidence}) and (2) the confidence level (conf.) for each magnitude and redshift bin.}
    \label{ch4_fig_color-sersci_vis_cnn_zbins}
\end{center}
\end{figure*}

The other quantity we consider is called `confidence\_flag' (Table~\ref{ch4_tab_catalog_col}) in this work which is determined by comparing the distributions of both S\'ersic index and colour ($g-i$) in each magnitude and redshift bin shown in Fig.~\ref{ch4_fig_color-sersci_vis_cnn_zbins} to the distributions of the reference samples. The confidence scheme is listed in Table~\ref{ch4_tab_confidence_flag}. From the discussion in Section~\ref{ch4_sec_compgz}, CNN classifications for galaxies with $16\le{i}<18$ and $z<0.25$ have our higher confidence class -- `superior confidence’. In addition, they are the reference for the others. Note that this analysis is strongly based on a presumption that the machine has a better performance when classifying galaxies that are in a similar observed `condition' (e.g., distance, magnitude, size) to the training set. However, this notion may or may not be true, which needs more investigation to be confirmed, and will be the topic of a forthcoming paper.

\begin{figure*}{}
\begin{center}
\graphicspath{}
	\includegraphics[width=2.1\columnwidth]{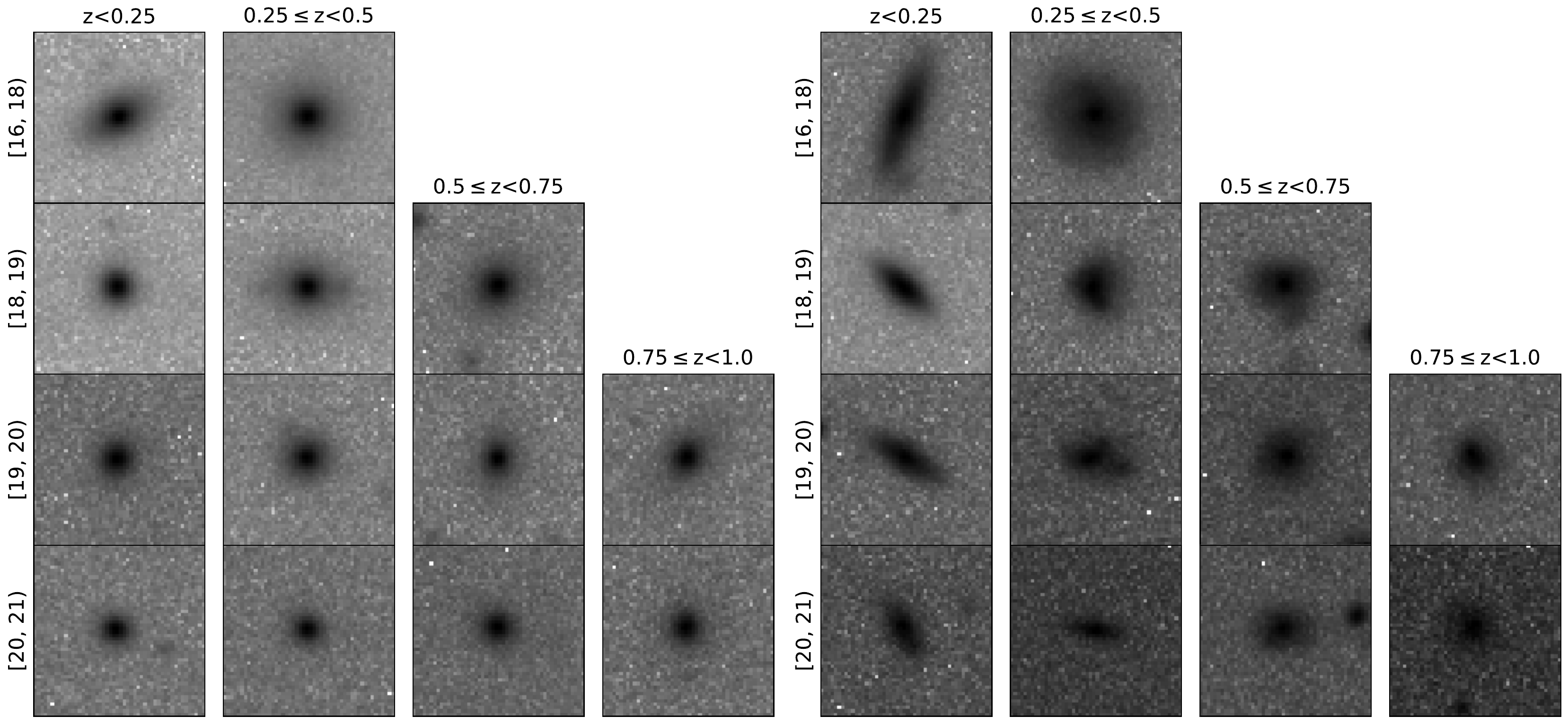}
   	\caption{Examples within different magnitude and redshift bins. CNN-classified Ellipticals are shown at the left side while the right side showcases Spirals (disk galaxies). Each row and column represents a range of magnitude and redshift, respectively.}
    \label{ch4_fig_example_in_magZbins}
\end{center}
\end{figure*}

In Fig.~\ref{ch4_fig_color-sersci_vis_cnn_zbins}, we further carry out statistical analyses to determine the confidence level for galaxies with $18\le{i}<21$ by subdividing the galaxies in each magnitude bin into 0.25-wide redshift bins. Galaxies are excluded from the catalogue if the number of galaxies with a given morphology type falls below 30 in a given bin since we do not have the necessary statistics to assess their reliability. The excluded galaxies are generally at the highest redshifts in their magnitude bins. Examples within each magnitude and redshift bin are shown in Fig.~\ref{ch4_fig_example_in_magZbins}. Using our CNN classifications and S\'ersic index information, we present examples of probable classes of Ellipticals (left) and Spirals (right) within different magnitude and redshift bins. Each row and column shows a range of magnitude and redshift, respectively.

\begin{figure*}{}
\begin{center}
\graphicspath{}
	\includegraphics[width=2.1\columnwidth]{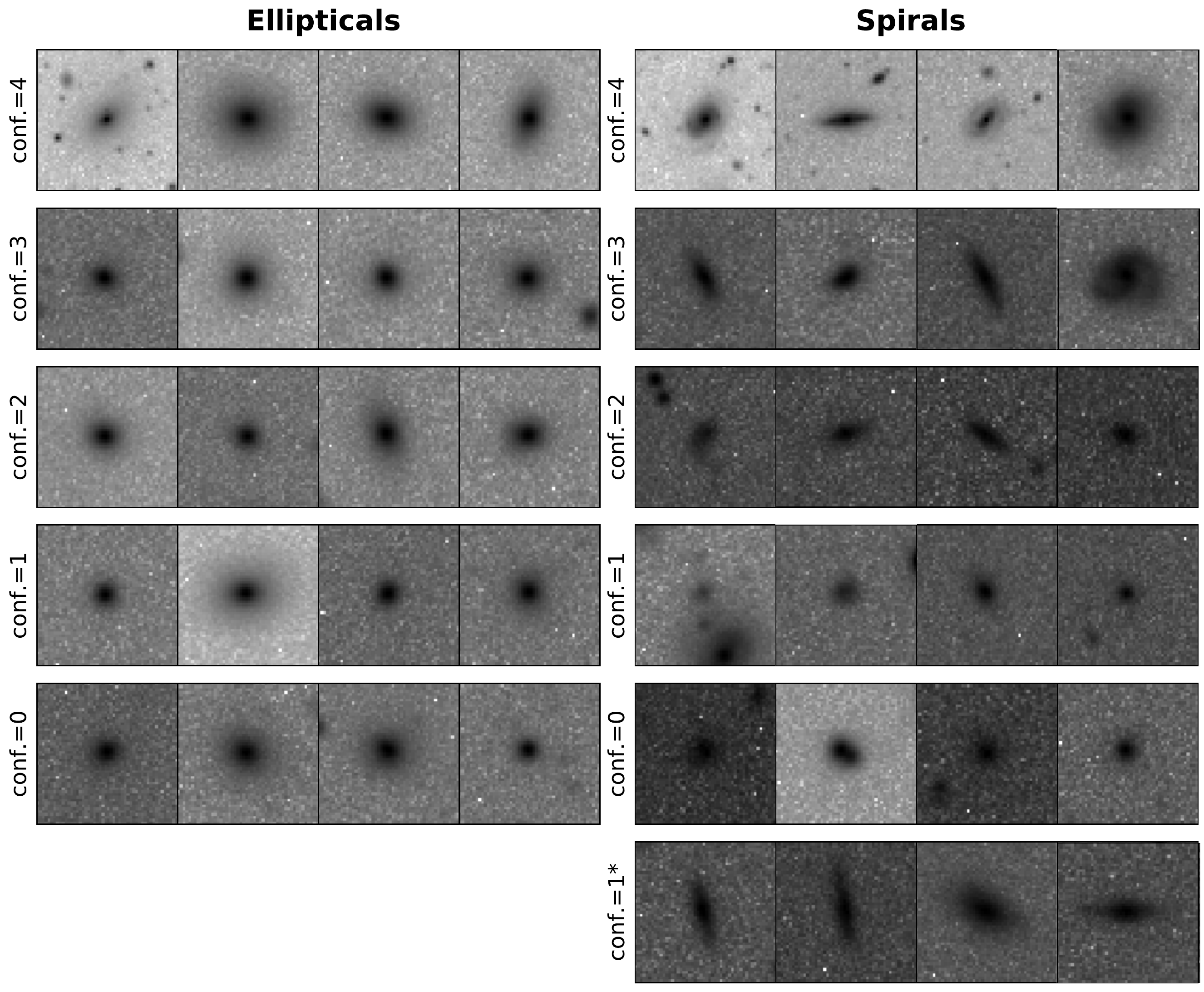}
   	\caption{Examples of each confidence level (see Table~\ref{ch4_tab_confidence_flag}) for the prediction of the two morphological types. The top row represents the most confident classification. CNN-classified Ellipticals are shown at the left side while the right panel presents CNN-classified Spirals. Confidence `1*' is for Spirals only as explained in Table~\ref{ch4_tab_confidence_flag}.}
    \label{ch4_fig_example_in_confidnece}
\end{center}
\end{figure*}

Each row in Fig.~\ref{ch4_fig_color-sersci_vis_cnn_zbins} shows the diagrams within a given magnitude range, while each column presents them in a different redshift bin. We use the `superior confidence’ classifications, top-left diagram, as reference to assess the confidence level of other ranges, i.e., the closer the distribution is to the reference, a higher value in the confidence scheme is assigned. We note that colour may actually not be a good criteria at the higher redshifts for the reasons described above. However, we use colour as it is one of the criteria that separates galaxy types quite cleanly at the lowest redshifts for high mass galaxies. What we require is specifically includes: (1) a clear distinction in both quantities between the two galaxy types; (2) the peaks of the distribution must be at similar locations for both morphologies when comparing with the reference, i.e., the S\'ersic index distributions should peak between 1 and 2 for Spirals and $\sim4$ for Ellipticals; (3) the median values of the S\'ersic index for both types are similar to the one in the reference within 1 $\sigma$ (median absolute deviation); and (4) no unusual features should be apparent in any of the single distributions (e.g., no bimodal or messy distributions). In Table~\ref{ch4_tab_confidence_flag}, a `high confidence' is assigned when all four criteria are satisfied. A `confidence' and a `less confidence' label is given when one or two of the criteria are missing, respectively. When more than three criteria are not satisfied, `no confidence' is allocated to the classifications for the galaxies within the corresponding magnitude and redshift ranges. Examples of the CNN classification with different confidence levels are shown in Fig.~\ref{ch4_fig_example_in_confidnece}.


\subsubsection{Magnitude bins: $16\le{i}<18$}
\label{ch4_sec_16to18}

In Section~\ref{ch4_sec_compgz}, we established the excellent performance of our CNN predictions for galaxies in the same magnitude and redshift ranges as the training set ($16\le{i}<18$ and $z<0.25$). On the first column of the first row in Fig.~\ref{ch4_fig_color-sersci_vis_cnn_zbins}, we show this robust conclusion again using a parametric morphology indicator, the S\'ersic index, and a generic galaxy property -- its colour. The distributions of both in this range are used as reference to determine the confidence level of other ranges. The median value of S\'ersic index for Spirals and Ellipticals is $1.61\pm0.60$ and $3.75\pm0.92$, respectively, in this magnitude and redshift range.

We notice that the {\it frac\_n} is relatively small in this region. In section~\ref{ch4_sec_compvis}, we discussed that our CNN classifies the class of Spirals mainly based on the presence of disk structure. Therefore, the small value of {\it frac\_n} in this region is due to the constraint on S\'ersic index in Equation~\ref{ch4_eq:sersic_confidence}. The fraction of CNN-classified Spirals with S\'ersic index between 2.5 and 4 is similar to the fraction of Ellipticals in this magnitude and redshift range. This indicates that the class of lenticular galaxies which is not well defined in our training set and has ambiguous structure could possibly confuse our CNN classifier \citep{Cheng2021}.

Next, we extend this examination to higher redshift but remain within the same magnitude range (second column at the first row in Fig.~\ref{ch4_fig_color-sersci_vis_cnn_zbins}). A clear distinction between two CNN predicted types in the S\'ersic index distribution can be seen within this redshift range, $0.25\le{i}<0.5$, and the peaks of both types are located in a sensible region. However, the CNN-classified Spirals have a broader distribution compared with the reference sample such that the median value is 2.66. Additionally, their colour distribution shows an apparent overlap with the CNN-classified Ellipticals. This suggests two possibilities: (1) our CNN classifier is being less accurate within this range, and/or (2) there are a fair number of galaxies with the structural features of Spirals but which are red in colour, particularly within $g-i$. Overall, we label the CNN predictions within this range as `less confidence'.

\subsubsection{Magnitude bins: $18\le{i}<19$}
\label{ch4_sec_18to19}

In the magnitude range $18\le{i}<19$, we have three redshift bins which include more than 30 galaxies with morphological measurements within each type: $z<0.25$, $0.25\le{z}<0.5$, and $0.5\le{z}<0.75$. In the first plot, we notice a good differentiation between the features of Ellipticals and Spirals such that the median value of the S\'ersic index for Spirals and Ellipticals is 1.32 and 3.18, respectively. However, the peak of the S\'ersic index distribution for Ellipticals is not located at $\sim4$. Therefore, we label the predictions of this range as those with `confidence'.

The second diagram ($0.25\le{z}<0.5$) has a slightly broader distribution of S\'ersic indices for CNN-classified Spirals with a median value of 2.25 compared to the reference. Except for this, a distinguishable separation in S\'ersic index distribution and colour distribution is presented. Hence, we recognise the CNN classified labels in this range as `confidence'.

Finally, the last panel shows an apparently overlapping colour distribution between Ellipticals and Spirals. In addition, a slightly bimodal structure is presented in the colour distribution. However, the median value of the S\'ersic indices for Spirals and Ellipticals is 1.88 and 3.52, respectively, which is acceptable when comparing with the values of the reference. However, the peak of S\'ersic index distribution for Spirals is off compared with the reference, i.e., not located between 1 and 2. We thus conservately label the classifications in this range as `no confidence'.  However, we cannot discount that these classifications are reliable given the range and distributions in S\'ersic indices.

\subsubsection{Magnitude bins: $19\le{i}<20$}
\label{ch4_sec_19to20}

In this fainter magnitude range, we observe an interesting result: a good consistency for our CNN predictions compared to the reference is found in the two higher redshifts bins, $0.25\le{z}<0.5$ and $0.5\le{z}<0.75$, than in the lower one. In these ranges, the median values of S\'ersic indices for each morphology type are within the ranges provided by the reference, and the peaks of the S\'ersic index distributions are reasonable. However, for the category of galaxies with redshifts $0.5\le{z}<0.75$, a clearly bimodal distribution is presented. We therefore give the morphological classifications for galaxies in these redshift ranges a `high confidence' and `confidence' label, respectively.

The low redshift interval ($z<0.25$; first column) shows a worse performance. We find a flat S\'ersic index distribution for the CNN-classified Ellipticals which peaks at roughly $n\sim2$ with a median value of 2.43. Additionally, although there is a separation in the colour distribution between the two types, the CNN-classified Ellipticals show a bimodal colour distribution which partially overlaps with the CNN-classified Spirals. Although the performance for Ellipticals in this redshift range is clearly worse, the behaviour for Spirals is significantly better: there is a fairly good discrimination in both the S\'ersic index and the colour distributions. This means that in this redshift range, our CNN-classified spiral sample has a high purity but not a high completeness. We therefore label the classifications made in this range as `less confidence' but with a `*' mark (Table~\ref{ch4_tab_confidence_flag}). The `*' indicates that this confidence level is only defined for CNN-classified Spirals, and the classified Ellipticals are labelled as `no confidence’. Clearly this sample cannot be used to find all Spirals, but we do have some confidence in the morphologies for the ones it does classify. In addition, there are a much larger number of CNN-classified Spirals than Ellipticals; therefore, the high {\it frac\_n} in this region supports the confidence assignment.

It seems counter-intuitive that a better performance is found for higher redshift galaxies than for lower redshift ones at these faint magnitudes. However, the reason is that the fainter galaxies in the training set tend to be low luminosity galaxies or are systems at higher redshifts. Therefore, there is a somewhat better overlap in the properties of faint higher redshift galaxies than there is for faint lower-redshift ones between the general DES Y3 sample and the training set. This issue is also discussed in computational science as an interesting issue called `The Elephant in the Room \citep{Rosenfeld2018}'. They proposed one of the reasons for this situation - `{\it Out of Distribution Examples}'. In our case, we interpret the distribution presented in faint lower-redshift galaxies as less likely to occur under our distributions of training sets.

Finally, for this magnitude range we give a `no confidence' label to the highest redshift range ($0.75\le{z}< 1.0$). This is due to the messy galaxy property distributions reflected in the bimodal colour distributions for both morphological types, a significantly higher S\'ersic index than expected for the CNN-classified Ellipticals, and a relatively low S\'ersic index for the CNN-classified Spirals. Interestingly, despite the relatively anomalous S\'ersic index, a fairly sharp differentiation between both types is shown in the S\'ersic index distributions. For the CNN-classified Ellipticals, this suggests a class of red galaxies which has a higher concentration and a more peaked surface brightness distribution than expected. This is an interesting conclusion from our CNN classification analysis that deserves to be explored further in future work.

\subsubsection{Magnitude bins: $20\le{i}<21$}
\label{ch4_sec_20to21}

As we get to fainter magnitudes using our CNN methodology to classify galaxies becomes more of a challenge. From Fig.~\ref{ch4_fig_roc_cm_vis_cnn} and Fig.~\ref{ch4_fig_color-sersci_vis_cnn}, we notice that there are also significantly fewer galaxies classified as Ellipticals by our CNN set up in this range, such that the CNN-classified E/S ratio is $\sim$0.0030 for total samples and $\sim$0.0056 for overlapping samples with morphological measurements, while the ones in other brighter ranges have a ratio over 0.1. This indicates that the bias self-correction by our CNN classifier might be overdone in this range compared to the brighter ranges. However, unlike the result shown in the range $21\le{i}\le21.5$ in Fig.~\ref{ch4_fig_color-sersci_vis_cnn}, a better and clearer separation between both types in S\'ersic index and colour is presented. Hence, we carry out a further investigation within different redshift bins for this range.

In the first plot on the bottom row in Fig.~\ref{ch4_fig_color-sersci_vis_cnn_zbins}, the distributions of CNN-classified Spirals are fairly reasonable with a median value of 1.04. However, a differing peak assignment of the S\'ersic index occurs within the CNN-classified Ellipticals. Therefore, we decided to assign a class of `less confidence' with `*' for this range, where `*' means this confidence label is for the Spirals (no confidence to the classification of Ellipticals). The {\it frac\_n} reflects the same support as the plot above within $19\le{i}<20$.

The second plot for this magnitude range in Fig.~\ref{ch4_fig_color-sersci_vis_cnn_zbins} shows a good separation between the two types of galaxies in S\'ersic index and colour space. Although a strong imbalance in the number of Ellipticals and Spirals still exists here, the differentiation in two types proves a certain degree of confidence to our CNN predictions. The median value of the S\'ersic index for Spirals and Ellipticals is 1.22 and 3.31, respectively, in this redshift range. However, the peak of the S\'ersic index distribution for the Ellipticals is off compared with the reference. Hence, we label the predictions in this range as `confidence'. The reason for this good separation in this significantly fainter magnitude range is also due to the effect discussed in Section~\ref{ch4_sec_19to20} that the galaxies in this magnitude ($20\le{i}<21$) and redshift ($0.25\le{z}<0.5$) range have similar galaxy features and galaxy properties to the reference samples. The shift in magnitude for these galaxies is due to the change in redshifts.

This situation is also demonstrated within the third plot of Fig.~\ref{ch4_fig_color-sersci_vis_cnn_zbins} ($20\le{i}<21$ and $0.5\le{z}<0.75$) whereby both types are distinguished in S\'ersic index distribution with a median value of 1.78 and 3.77 for Spirals and Ellipticals, respectively. However, CNN-classified Spirals have a relatively flat colour distribution which shows an indication of a small bimodal distribution and prevents a clear separation. Hence, a class of `less confidence' is assigned to this range. Finally, the last diagram shows a messy distribution. Therefore, we simply label this range as a `no confidence' class.

\begin{table}
	\centering
	\begin{tabular}{clc}
		\hline
		\multicolumn{1}{l}{labels} & {representation} & {number of galaxies}\\
		\hline\hline
		\multicolumn{1}{l}{4} & {superior confidence} & {672,927}\\
		\multicolumn{1}{l}{3} & {high confidence} & {3,409,459}\\
		\multicolumn{1}{l}{2} & {confidence} & {9,230,182}\\
		\multicolumn{1}{l}{1} & {less confidence} & {4,347,472}\\
		\multicolumn{1}{l}{1*} & {less confidence} & {2,599,656}\\
		\multicolumn{1}{l}{} & {(for Spirals only)} & {}\\
		\multicolumn{1}{l}{0} & {no confidence} & {859,411}\\
		\hline\hline
		\multicolumn{1}{l}{Total} & {} & {21,119,107}\\
		\hline
	\end{tabular}
	\caption{Content of the {\it confidence\_flag} (column 7) shown in Table~\ref{ch4_tab_catalog_col}. The 'superior confidence' flag is for classifications within the same magnitude and redshift ranges as the training set. The details of other levels are described in Section~\ref{ch4_sec_confidence_level}. The total number of classification provided in this catalogue is 21,119,107.}
	\label{ch4_tab_confidence_flag}
\end{table}

\subsection{VIPERS spectral classification}
\label{ch4_sec_viper_comp}

After Section~\ref{ch4_sec_compvis} and Section~\ref{ch4_sec_confidence_level}, we finalise the number of galaxy classifications in our final catalogue. In this Section, we compare our CNN predictions with the spectral classification from VIPERS presented in \citet[][Section~\ref{ch4_sec_vipers_cata}]{Siudek2018}. The number of overlapping samples between their spectral classification catalogue and our final catalogue is 10,254; of which, 9,459 galaxies have a high class membership probability in their catalogue.  This is enough galaxies to test how our classifications agree with those based on spectroscopy.

\begin{figure}{}
\begin{center}
\graphicspath{}
	\includegraphics[width=\columnwidth]{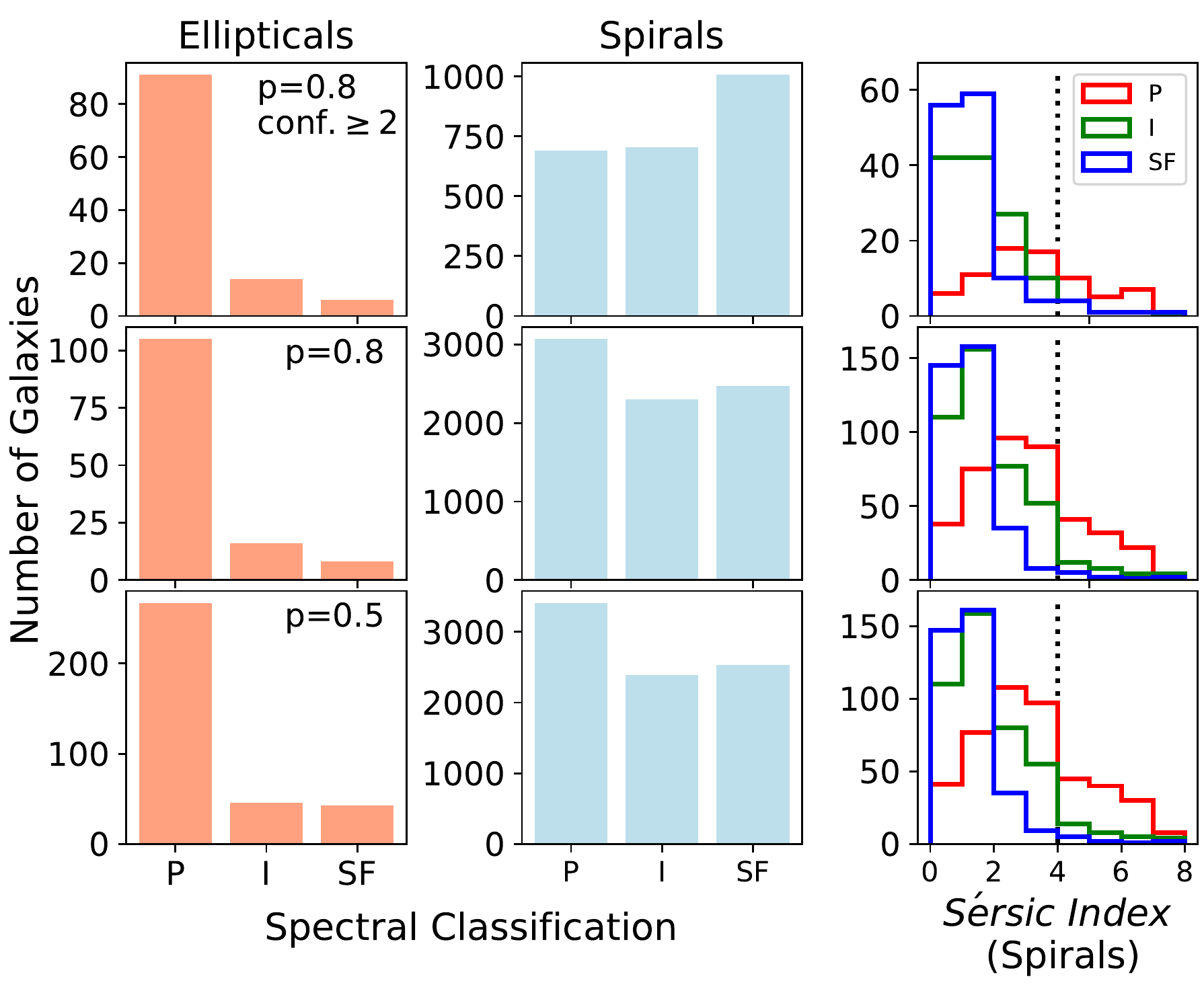}
   	\caption{Comparison between our CNN predictions and the VIPER unsupervised spectral classification. From the bottom to the top row, a probability threshold of 0.5, 0.8, and 0.8 with the confidence level greater than 2, is applied in each row. The P, I, SF at the $x$-axis of the first two panels represents the spectral classifications of passive, intermediate, and star-forming galaxies, respectively. The last panel shows the S\'ersic index distribution of Spirals labelled by our CNN. Different colours, red, green, blue, represents galaxies with different spectral classifications, passive (P), intermediate (I), star-forming (SF), respectively. The vertical dotted line indicates where the S\'ersic index equals to 4.}
    \label{ch4_fig_comp_spectral_cnn}
\end{center}
\end{figure}

Three main classes of spectral-types for galaxies  - passive (P), intermediate (I), star-forming (SF) are defined in this catalogue which we use to examine our CNN classifications. Fig.~\ref{ch4_fig_comp_spectral_cnn} shows that Ellipticals labelled by our CNN are mostly passive, such that we find fractions of 0.75, 0.81, and 0.82 are passive CNN-classified Ellipticals from $p=0.5$, $p=0.8$, to $p=0.8$ with conf. $\ge2$, but CNN-labelled Spirals show a mixture of three classes. We further examine the S\'ersic index distribution of the CNN-labelled Spirals in the last panel of Fig.~\ref{ch4_fig_comp_spectral_cnn}. This figure shows that the CNN-labelled Spirals at the passive and intermediate spectral stages are mostly disky, i.e., those galaxies with S\'ersic index distribution at $<4$. The fraction of intermediate CNN-labelled Spirals with S\'ersic indices smaller than 4 is 0.87, 0.88, and 0.93 ($<3$ is 0.75, 0.76, and 0.85) of the total sample in each row, from $p=0.5$, $p=0.8$, to $p=0.8$ with conf. $\ge2$, while the fraction of passive CNN-labelled Spirals is 0.60, 0.63, and 0.57 ($<3$ is 0.42, 0.44, and 0.38).

Since galaxies with fainter magnitudes and at higher redshift might not have a clear visual spiral arms (Fig.~\ref{ch4_fig_example_in_magZbins}), in addition to the possibility of passive Spirals \citep{Masters2010}, our CNN is likely to classify passive disk galaxies, such as lenticulars, into the class of Spirals.

\subsection{Non-parametric methods and galaxy properties}
\label{ch4_sec_nonparams}

Another examination is carried out using non-parametric methods such as the {\it CAS system} (Concentration, Asymmetry, and Smoothness/Clumpiness), Gini coefficient, and M20. In this study, the non-parametric measurements are from \citet{Tarsitano2018} using the $i$-band images, and we use the measurements after applying the selection criteria described in Section~\ref{ch4_sec_des_y1cata}.

\begin{figure*}{}
\begin{center}
\graphicspath{}
	\includegraphics[width=2.1\columnwidth]{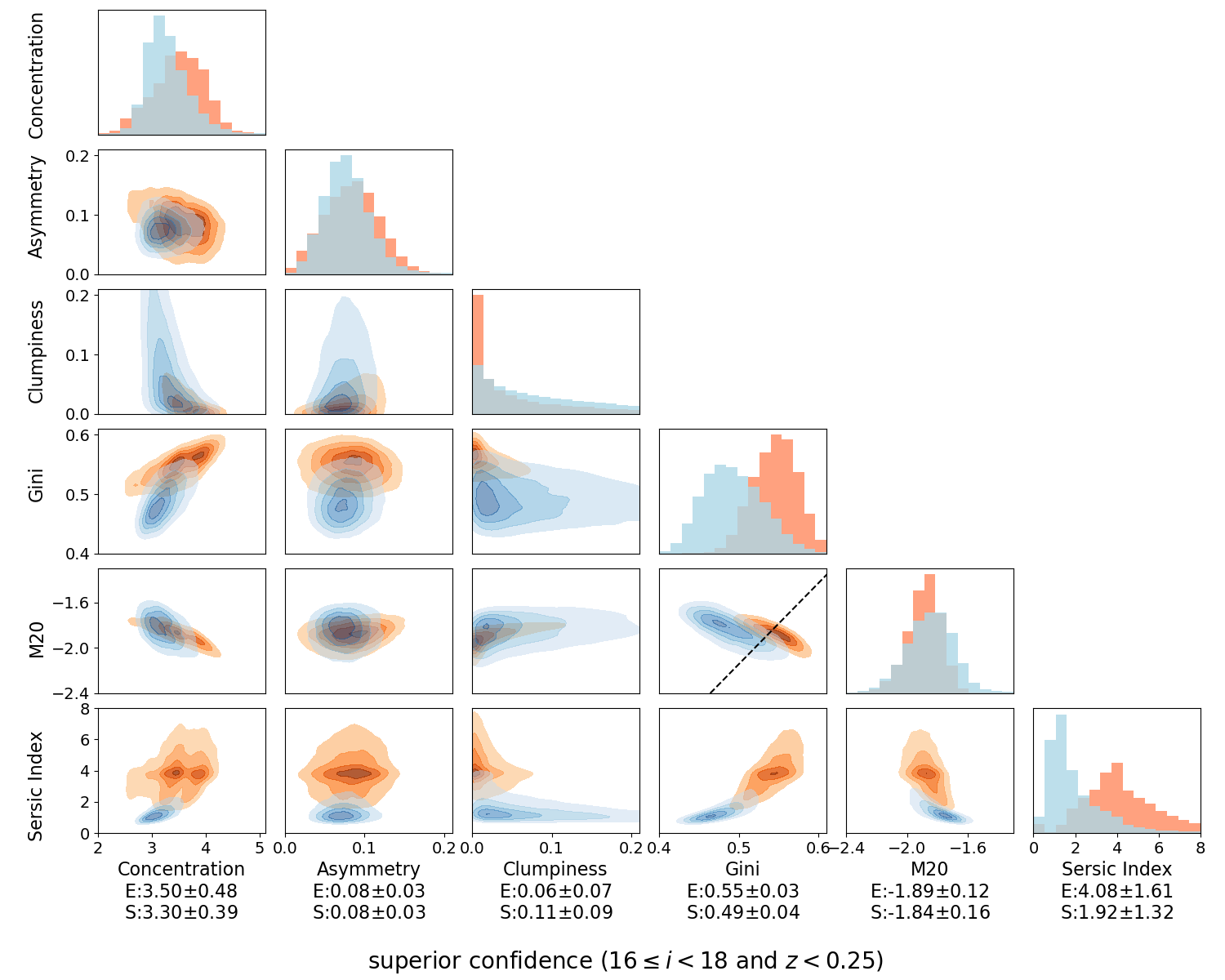}
   	\caption{The pair plots of six morphological parameters: concentration, asymmetry, clumpiness, Gini, M20, and S\'ersic index labelled by the CNN classifications with `superior confidence'. The colour shadings represent the CNN classifications. The red/orange and blue colour are for Ellipticals (E) and Spirals (S), respectively. The mean value of each parameter for both types with the standard deviation is shown below each column. The black dashed line shows a cut from \citet{Lotz2008} to separate Ellipticals and Spirals based on the M20 and the Gini coefficients.}
    \label{ch4_fig_superior_nonparams}
\end{center}
\end{figure*}

Furthermore, this validation can work in both directions. We can use the non-parametric measurements to check the robustness of our CNN-based morphological classifications, while also use our most reliable morphological classifications (those with `superior confidence') to assess the ability of non-parametric methods to separate the Ellipticals and the Spirals (Fig.~\ref{ch4_fig_superior_nonparams}). Such an analysis of non-parametric measurements as proxies for morphology has never before been carried out with samples as large as ours. In this work we include over 100,000 galaxies in the `superior confidence' category from our DES Y3 morphological classifications.

In Fig.~\ref{ch4_fig_superior_nonparams}, we show the pair plots of six different parameters: concentration (C), asymmetry (A), clumpiness (S), Gini, M20, and S\'ersic index. For the A, S parameters, we only showcase the data with values smaller than 0.2 to focus on `typical galaxies'. The S\'ersic index is used as a comparison to the non-parametric methods, and it is one of the main features used to define the confidence level (Section~\ref{ch4_sec_confidence_level}).  It shows a clear separation between the two morphological types here. In addition to this, we note that only the Gini coefficient shows a consistently distinguished difference between the two types in the histogram.

The Gini coefficient (G) reflects the inequality of the flux distributed among the pixels of a given galaxy; if $G=1$, the light is concentrated in one pixel, conversely, $G=0$ means that the light is uniformly distributed to every pixel. Therefore, the Gini coefficient is somewhat analogous to the concept of concentration, and Ellipticals generally have a higher value than Spirals. Nevertheless, the concentration does not show a separation as good as the one for the Gini coefficient. A slight shift between the peaks of the two morphological types is shown in the histogram of the concentration; however, a large overlapping area is also shown. Additionally, the difference of the mean concentration values between both types is relatively small compared with previous studies \citep{Conselice2003, Hernandez-Toledo2008, Hambleton2011}. On the other hand, both asymmetry and clumpiness also fail to show a consistent distinction between the two morphological types in our analysis.

Finally, the M20 histogram does not show a clear separation between the two morphological types either. However, a clean separation does shows itself in the contour of the Gini coefficient and M20. The black dashed line indicates a cut used to separate Ellipticals and Spirals and described in \citet{Lotz2008} such that

\begin{equation}
    G=0.14{ M }_{ 20 }+0.8.
	\label{ch4_eq:Gini_M20}
\end{equation}

Thus we find that the Gini coefficient is a possible better tracer of the overall structure of a galaxy than any other non-parametric morphological quantities such as C, A, S, and M20  \citep{Zamojski2007} when separating Ellipticals from Spirals. 

\section{Galaxy Morphological Classification Catalogue}
\label{ch4_sec_y3catalog}

In this paper, with the convolutional neural network (CNN) trained with the subset of the DES Y1 data with the GZ1 labels corrected in C20 (Section~\ref{ch4_sec_traindata} and Section~\ref{ch4_sec_gz_cata}), we provide one of the largest catalogue to date with galaxy morphological classifications for over 20 million galaxies from the DES Y3 data ($16\le{i}<21$ and $z<1.0$; Section~\ref{ch4_sec_y3data}; along with the companion catalogue produced by \citet{Vega-Ferrero2020}). As mentioned in Section~\ref{ch4_sec_intro}, an extensive comparison of the these two catalogues is ongoing, the result of which will be published in a future paper.

The items provided in our catalogue of morphological types are listed in Table~\ref{ch4_tab_catalog_col}. The average predicted probabilities from the five individual CNN models (Section~\ref{ch4_sec_cnn}) are used as the final probabilities of being Ellipticals (pE) and Spirals (pS). With this quantity, users can apply a probability threshold, which determines the tolerance of the accuracy of the morphological classification, to fit with their scientific goals. In this catalogue, we provide the classification label based on a threshold of 0.8 ({\it MORPH\_FLAG}) for the user's convenience.

\begin{table}
	\centering
	\begin{tabular}{cll}
		\hline
		\multicolumn{1}{l}{Col.} & {Keyword} & {Description}\\
		\hline\hline
		\multicolumn{1}{l}{1} & {DES\_Y3\_ID} & {DES Y3 ID} \\
		\multicolumn{1}{l}{2} & {RA} & {right ascension} \\
		\multicolumn{1}{l}{3} & {DEC} & {declination} \\
		\multicolumn{1}{l}{4} & {pE} & {probability of being Ellipticals} \\
		\multicolumn{1}{l}{5} & {pS} & {probability of being Spirals} \\
		\multicolumn{1}{l}{6} & {MORPH\_FLAG} & {CNN predictions with a} \\
		\multicolumn{1}{l}{} & {} & {probability threshold of 0.8} \\
		\multicolumn{1}{l}{7} & {confidence\_flag} & {confidence level for predictions} \\
		\multicolumn{1}{l}{8} & {frac\_n} & {Fraction of predictions which} \\
		\multicolumn{1}{l}{} & {} & {satisfy S\'ersic index criteria} \\
		\multicolumn{1}{l}{9} & {MAG\_I} & {$i$-band magnitude.} \\
		\multicolumn{1}{l}{} & {} & {{\it (MAG\_AUTO\_I)}} \\
		\multicolumn{1}{l}{10} & {ZMEAN} & {photometric redshift.} \\
		\multicolumn{1}{l}{} & {} & {{\it (DNF\_ZMEAN\_MOF)}} \\
		\hline
	\end{tabular}
	\caption{Content of the catalogue published with this paper. Columns 9 to 10 are quantities which are taken directly from the DES Y3 GOLD catalogue, and the corresponding column names are highlighted and placed within brackets in the description.}
	\label{ch4_tab_catalog_col}
\end{table}


Our CNN classifier is trained with bright galaxies (magnitudes $16\le{i}<18$) at low redshift ($z<0.25$). Therefore, for users who are more comfortable with our machine's predictions when applied to galaxies with similar condition to the training set, we carried out a statistical analysis (Section~\ref{ch4_sec_confidence_level}) to investigate the impact when the target data has a worse image quality than the training set due to faintness and redshift effects. Within this analysis, we provide a confidence flag for every galaxy (Table~\ref{ch4_tab_confidence_flag}) within our CNN classification final catalogue. In addition, another flag, {\it frac\_n}, which is defined as the fraction of predictions that satisfy the S\'ersic index criteria (Eqation~\ref{ch4_eq:sersic_confidence}), serves a similar purpose. Overall, over 20 million CNN classifications with an assigned confidence level are included in our final catalogue; of which, $\sim$670,000 galaxies have a `superior confidence', $\sim$3.4 millions of galaxies are assigned as a `high confidence' classification, and $\sim$9 million galaxies have a `confidence' label. Finally, in columns 9-10 in Table~\ref{ch4_tab_catalog_col} we provide magnitude and redshift information directly from the DES Y3 GOLD catalogue to allow customised magnitude/redshift cut when applying our predictions.

\section{Summary}
\label{ch4_sec_conclusion}

We present in this paper one of the largest galaxy morphological classification catalogue produced to date (along with the other DES catalogue presented in \citet{Vega-Ferrero2020}, using the Dark Energy Survey (DES) Y3 data with over 20 million galaxies. We carry out these classifications using convolutional neural networks (CNN) trained with the subset of a DES Y1 data. The {\bf corrected} debiased labels, which are initially from the Galaxy Zoo 1 (GZ1) catalogue and corrected in C20, are used to label our training set (Section~\ref{ch4_sec_traindata}). With a combination of three different types of input, including: linear images, log images, and HOG images (Section~\ref{ch4_sec_preprocessing}), our CNN classifier reaches an accuracy of over 99\% when compared with the GZ1 classifications ($i$-band magnitude<18 and redshift<0.25). The majority of mismatches occurs in the case when a galaxy is classified as a Spirals by our CNN but as Ellipticals by GZ1. The reason behind this mismatch is likely to be the better resolution and deeper depth of the DES imaging data which reveals unnoticeable structure in the data used in GZ1 from the Sloan Digital Sky Survey \citep[more discussion in][]{Cheng2020a}. Additionally, training with the {\bf corrected} debiased labels, our CNN classifier is shown to be self-debiased and more accurate in classifying disk galaxies which human visual classifications have difficulty detecting at faint magnitudes down to $i\sim21$ (see Section~\ref{ch4_sec_compvis}).

Trained with bright galaxies at low redshift, our CNN classifier is statistically assessed for its performance when used to predict morphologies for fainter galaxies at higher redshift. This assessment provides an investigation about how well a machine trained within one domain can be applied to the conditions in different domains; in our case, we applied the machine trained with bright galaxies ($i<18$) at low redshift ($z<0.25$) to a fainter galaxies ($i\ge18$). Using a cross-validation with the S\'ersic index and galaxy colour ($g-i$), we provide a confidence evaluation scheme to our CNN classifications (Table~\ref{ch4_tab_confidence_flag}) through a statistical analysis of data in different magnitude and redshift bins (Section~\ref{ch4_sec_confidence_level}). We define six confidence levels by comparing with the S\'ersic indices and colour distributions of the data within the same coverage as the training set. In this assessment, we find that a better confidence is assigned to faint galaxies at higher redshift compared to galaxies with fainter magnitudes, but at lower redshift. For example, the confidence of predictions for galaxies with $19\le{i}<20$ at $0.25\le{z}<0.5$ is higher than the one at $z<0.25$ at the same magnitude range. Faint galaxies in the training set are generally at higher redshift. A faint galaxy at relatively low redshift is an anomaly, in the sense that they do not existed in our training domain, in the machine's view. Thus, the machine gives a better prediction for fainter galaxies at higher redshift than systems at lower redshift, even though the magnitude and redshift ranges of these galaxies are beyond the ranges of the training set. Finally, we conclude that over 13 million galaxies (over 60\% of the total classifications) have at least a `confidence' level as defined in our work.

As a part of the validation, we carry out a large examination of non-parametric methods such as the {\it CAS system} (Concentration, Asymmetry, and Smoothness/Clumpiness), the Gini coefficient, and M20 using over 100,000 classifications with structural measurements from \citet{Tarsitano2018}. From this we conclude that the Gini coefficient shows the most significant distinction, as a single parameter, between Ellipticals and Spirals within all parameters tested.  Additionally, with a combination of the M20 index, a straight line \citep{Lotz2008} can be drawn to separate these two types (Fig.~\ref{ch4_fig_superior_nonparams}).

In addition, we compare our CNN predictions with spectral classification from VIPERS presented in \citet{Siudek2018}. The result shows that the CNN-classified Ellipticals are mostly passive with a passive fraction of over 0.75. On the other hand, the CNN-classified Spirals show a mixture of passive, intermediate, and star-forming classes, but the majority have disk-like structures (S\'ersic index$<3$). In addition to the possibility of passive Spirals, lenticulars are also responsible for the fraction of passive CNN-labelled Spirals in our case.

In this work, we used only observed data (bright galaxies at low redshift) to train our CNN which limits potential applications to very faint galaxies. However, through the analysis carried out in this work, we notice that our machine classifies disky galaxies with round and blurred structure to the class of Spirals, while humans usually misclassify these systems as Ellipticals (Section~\ref{ch4_sec_compvis}). This supports the usefulness of our machine classification for fainter galaxies. Users can straightly utilise the predicted probabilities ({\it pE} and {\it pS} in Table~\ref{ch4_tab_catalog_col}) to obtain galaxy morphology predictions. The {\it MORPH\_FLAG} provided in Table~\ref{ch4_tab_catalog_col} uses a probability threshold of 0.8 to define Spirals (1) and Ellipticals (0) for users' convenience.

Our new morphological catalogue allows a variety of new approaches towards understanding galaxy properties and evolution that involve morphology that could not be carried out before. For example, non-parametric analysis methods of galaxy structure can be assessed using an unprecedented sample not only in size but also in quality. Our catalogue can also be used to cross-validate other classification methods, and to explore galaxy properties and environment as a function of morphology with superb statistics.   Future papers will examine these feature of the galaxy population and galaxy evolution with morphology using our classifications.

Scientifically, there are of course a myriad of other uses for our catalogue, as morphology is one of the fundamental properties of galaxies.  For the time being this will remain one of the largest set of morphological classifications available for analysis for any survey done to date (along with the companion classification catalogue produced by \citet{Vega-Ferrero2020}). Our methodology is also scalable and meant to be of use for applications to future imaging data sets such as the ones that will eventually be created from the Euclid Space Telescope and the Vera Rubin Observatory Legacy Survey of Space and Time, among others.

\section*{Acknowledgements}
T.-Y. Cheng acknowledges the support of the Vice-Chancellor's Scholarship from the University of Nottingham and STFC grants ST/T000244/1 and ST/P000541/1 at the Durham University. Funding for the DES Projects has been provided by the U.S. Department of Energy, the U.S. National Science Foundation, the Ministry of Science and Education of Spain, the Science and Technology Facilities Council of the United Kingdom, the Higher Education Funding Council for England, the National Center for Supercomputing Applications at the University of Illinois at Urbana-Champaign, the Kavli Institute of Cosmological Physics at the University of Chicago, the Center for Cosmology and Astro-Particle Physics at the Ohio State University, the Mitchell Institute for Fundamental Physics and Astronomy at Texas A$\&$M University, Financiadora de Estudos e Projetos, Fundação Carlos Chagas Filho de Amparo à Pesquisa do Estado do Rio de Janeiro, Conselho Nacional de Desenvolvimento Científico e Tecnológico and the Ministério da Ciência, Tecnologia e Inovação, the Deutsche Forschungsgemeinschaft, and the Collaborating Institutions in the Dark Energy Survey.

The Collaborating Institutions are Argonne National Laboratory, the University of California at Santa Cruz, the University of Cambridge, Centro de Investigaciones Energéticas, Medioambientales y Tecnológicas-Madrid, the University of Chicago, University College London, the DES-Brazil Consortium, the University of Edinburgh, the Eidgenössische Technische Hochschule (ETH) Zürich, Fermi National Accelerator Laboratory, the University of Illinois at Urbana-Champaign, the Institut de Ciències de l'Espai (IEEC/CSIC), the Institut de Física d'Altes Energies, Lawrence Berkeley National Laboratory, the Ludwig-Maximilians Universität München and the associated Excellence Cluster Universe, the University of Michigan, the National Optical Astronomy Observatory, the University of Nottingham, The Ohio State University, the University of Pennsylvania, the University of Portsmouth, SLAC National Accelerator Laboratory, Stanford University, the University of Sussex, Texas A$\&$M University, and the OzDES Membership Consortium.

Based in part on observations at Cerro Tololo Inter-American Observatory, National Optical Astronomy Observatory, which is operated by the Association of Universities for Research in Astronomy (AURA) under a cooperative agreement with the National Science Foundation.

The DES data management system is supported by the National Science Foundation under grant numbers AST-1138766 and AST-1536171. The DES participants from Spanish institutions are partially supported by MINECO under grants AYA2015-71825, ESP2015-66861, FPA2015-68048, SEV-2016-0588, SEV-2016-0597, and MDM-2015-0509, some of which include ERDF funds from the European Union. IFAE is partially funded by the CERCA programme of the Generalitat de Catalunya. Research leading to these results has received funding from the European Research Council under the European Union's Seventh Framework Program (FP7/2007-2013) including ERC grant agreements 240672, 291329, and 306478. We acknowledge support from the Australian Research Council Centre of Excellence for All-sky Astrophysics (CAASTRO), through project number CE110001020, and the Brazilian Instituto Nacional de Ciencia e Tecnologia (INCT) e-Universe (CNPq grant 465376/2014-2).


\section*{Data Availability}
This DES Y3 morphological classification catalogue is currently not publicly available but can be shared on request to the corresponding author. It will be available soon in the Dark Energy Survey Data Management (DESDM) system.

\bibliographystyle{mnras}
\bibliography{ms}

\appendix
\section*{Affiliations}
$^{1}$Centre of Extragalactic Astronomy, Durham University, Stockton Rd, Durham DH1 3LE, UK\\
$^{2}$School of Physics and Astronomy, University of Nottingham, University Park, Nottingham, NG7 2RD, UK\\
$^{3}$Jodrell Bank Centre for Astrophysics, University of Manchester, Oxford Road, Manchester UK \\
$^{4}$Departamento de F\'isica Matem\'atica, Instituto de F\'isica, Universidade de S\~ao Paulo, CP 66318, S\~ao Paulo, SP, 05314-970, Brazil \\
$^{5}$Laborat\'orio Interinstitucional de e-Astronomia - LIneA, Rua Gal. Jos\'e Cristino 77, Rio de Janeiro, RJ - 20921-400, Brazil \\
$^{6}$Fermi National Accelerator Laboratory, P. O. Box 500, Batavia, IL 60510, USA \\
$^{7}$Instituto de F\'{i}sica Te\'orica, Universidade Estadual Paulista, S\~ao Paulo, Brazil \\
$^{8}$Cavendish Laboratory Astrophysics Group, University of Cambridge, Madingley Road, Cambridge CB3 0HA, UK \\
$^{9}$Kavli Institute for Cosmology, University of Cambridge, Madingley Road, Cambridge CB3 0HA, UK \\
$^{10}$Department of Physics \& Astronomy, University College London, Gower Street, London, WC1E 6BT, UK \\
$^{11}$Kavli Institute for Particle Astrophysics \& Cosmology, P. O. Box 2450, Stanford University, Stanford, CA 94305, USA \\
$^{12}$SLAC National Accelerator Laboratory, Menlo Park, CA 94025, USA \\
$^{13}$Center for Astrophysical Surveys, National Center for Supercomputing Applications, 1205 West Clark St., Urbana, IL 61801, USA \\
$^{14}$Department of Astronomy, University of Illinois at Urbana-Champaign, 1002 W. Green Street, Urbana, IL 61801, USA \\
$^{15}$Institut de F\'{\i}sica d'Altes Energies (IFAE), The Barcelona Institute of Science and Technology, Campus UAB, 08193 Bellaterra (Barcelona) Spain \\
$^{16}$Center for Cosmology and Astro-Particle Physics, The Ohio State University, Columbus, OH 43210, USA \\
$^{17}$Astronomy Unit, Department of Physics, University of Trieste, via Tiepolo 11, I-34131 Trieste, Italy \\
$^{18}$INAF-Osservatorio Astronomico di Trieste, via G. B. Tiepolo 11, I-34143 Trieste, Italy \\
$^{19}$Institute for Fundamental Physics of the Universe, Via Beirut 2, 34014 Trieste, Italy \\
$^{20}$Observat\'orio Nacional, Rua Gal. Jos\'e Cristino 77, Rio de Janeiro, RJ - 20921-400, Brazil \\
$^{21}$Department of Physics, University of Michigan, Ann Arbor, MI 48109, USA \\
$^{22}$Centro de Investigaciones Energ\'eticas, Medioambientales y Tecnol\'ogicas (CIEMAT), Madrid, Spain \\
$^{23}$Department of Astronomy and Astrophysics, University of Chicago, Chicago, IL 60637, USA \\
$^{24}$Kavli Institute for Cosmological Physics, University of Chicago, Chicago, IL 60637, USA \\
$^{25}$Department of Physics and Astronomy, University of Pennsylvania, Philadelphia, PA 19104, USA \\
$^{26}$Santa Cruz Institute for Particle Physics, Santa Cruz, CA 95064, USA \\
$^{27}$Department of Astronomy, University of Michigan, Ann Arbor, MI 48109, USA \\
$^{28}$Institute of Theoretical Astrophysics, University of Oslo. P.O. Box 1029 Blindern, NO-0315 Oslo, Norway \\
$^{29}$Institut d'Estudis Espacials de Catalunya (IEEC), 08034 Barcelona, Spain \\
$^{30}$Institute of Space Sciences (ICE, CSIC),  Campus UAB, Carrer de Can Magrans, s/n,  08193 Barcelona, Spain \\
$^{31}$Instituto de Fisica Teorica UAM/CSIC, Universidad Autonoma de Madrid, 28049 Madrid, Spain \\
$^{32}$Institute of Astronomy, University of Cambridge, Madingley Road, Cambridge CB3 0HA, UK \\
$^{33}$Department of Physics, Stanford University, 382 Via Pueblo Mall, Stanford, CA 94305, USA \\
$^{34}$School of Mathematics and Physics, University of Queensland,  Brisbane, QLD 4072, Australia \\
$^{35}$Department of Physics, The Ohio State University, Columbus, OH 43210, USA \\
$^{36}$Center for Astrophysics $\vert$ Harvard \& Smithsonian, 60 Garden Street, Cambridge, MA 02138, USA \\
$^{37}$Department of Astronomy/Steward Observatory, University of Arizona, 933 North Cherry Avenue, Tucson, AZ 85721-0065, USA \\
$^{38}$Australian Astronomical Optics, Macquarie University, North Ryde, NSW 2113, Australia \\
$^{39}$Lowell Observatory, 1400 Mars Hill Rd, Flagstaff, AZ 86001, USA \\
$^{40}$Instituci\'o Catalana de Recerca i Estudis Avan\c{c}ats, E-08010 Barcelona, Spain \\
$^{41}$Physics Department, 2320 Chamberlin Hall, University of Wisconsin-Madison, 1150 University Avenue Madison, WI  53706-1390 \\
$^{42}$Department of Astrophysical Sciences, Princeton University, Peyton Hall, Princeton, NJ 08544, USA \\
$^{43}$School of Physics and Astronomy, University of Southampton,  Southampton, SO17 1BJ, UK \\
$^{44}$Computer Science and Mathematics Division, Oak Ridge National Laboratory, Oak Ridge, TN 37831 \\
$^{45}$Institute of Cosmology and Gravitation, University of Portsmouth, Portsmouth, PO1 3FX, UK \\

\label{lastpage}
\end{document}